\documentclass[a4paper,11pt]{article}

\usepackage{amsmath,amsthm,amsbsy,amssymb,amsfonts,graphics,ascmac,mathrsfs,bm,accents,cite}
\usepackage[dvipdfmx]{graphicx}

\makeatletter
\catcode`\@=11
\@addtoreset{equation}{section}

\makeatother

\renewcommand{\thefootnote}{\arabic{footnote}}
\makeatletter
\DeclareRobustCommand\widecheck[1]{{\mathpalette\@widecheck{#1}}}
\def\@widecheck#1#2{%
   \setbox\z@\hbox{\m@th$#1#2$}%
   \setbox\tw@\hbox{\m@th$#1%
      \widehat{%
         \vrule\@width\z@\@height\ht\z@
         \vrule\@height\z@\@width\wd\z@}$}%
   \dp\tw@-\ht\z@
   \@tempdima\ht\z@ \advance\@tempdima2\ht\tw@ \divide\@tempdima\thr@@
   \setbox\tw@\hbox{%
      \raise\@tempdima\hbox{\scalebox{1}[-1]{\lower\@tempdima\box\tw@}}}%
   {\ooalign{\box\tw@ \cr \box\z@}}}
\makeatother

\newcommand{\gs}{g_s}
\newcommand{\ls}{l_s}
\newcommand{\ap}{\alpha'}
\newcommand{\Exp}[1]{\operatorname{e}^{#1}}
\newcommand{\abs}[1]{{\lvert #1 \rvert}}
\newcommand{\absb}[1]{{\bigl\lvert #1 \bigr\rvert}}
\newcommand{\rmd}{{\mathrm{d}}}
\newcommand{\ii}{{\hspace{0.7pt}\mathrm{i}\hspace{0.7pt}}}

\newcommand{\im}{\operatorname{Im}}
\newcommand{\nn}{\nonumber}
\newcommand{\bpm}{\begin{pmatrix}}
\newcommand{\epm}{\end{pmatrix}}

\newcommand{\cA}{\mathcal A}
\newcommand{\cF}{\mathcal F}
\newcommand{\cH}{\mathcal H}
\newcommand{\cM}{\mathcal M}
\newcommand{\cQ}{\mathcal Q}
\newcommand{\cR}{\mathcal R}
\newcommand{\cT}{\mathcal T}

\newcommand{\lR}{{\mathbb R}}
\newcommand{\lZ}{{\mathbb Z}}

\newcommand{\sfm}{\mathsf{m}}
\newcommand{\sfn}{\mathsf{n}}
\newcommand{\sfg}{\mathsf{g}}
\newcommand{\sfR}{\mathsf{R}}
\newcommand{\sfA}{\mathsf{A}}

\newcommand{\DD}{\text{D}}
\newcommand{\EE}{\text{E}}
\newcommand{\FF}{\text{F}}
\newcommand{\GL}{\text{GL}}
\newcommand{\KK}{\text{KK}}
\newcommand{\NS}{\text{NS}}
\newcommand{\OO}{\text{O}}
\newcommand{\PP}{\text{P}}
\newcommand{\SL}{\text{SL}}
\newcommand{\WZ}{\text{WZ}}

\newcommand{\bfa}{\mathbf{a}}
\newcommand{\bfb}{\mathbf{b}}
\newcommand{\bfc}{\mathbf{c}}
\newcommand{\bfp}{\mathbf{p}}
\newcommand{\bfq}{\mathbf{q}}
\newcommand{\bfr}{\mathbf{r}}
\newcommand{\bfs}{\mathbf{s}}
\newcommand{\bft}{\mathbf{t}}
\newcommand{\II}{\mathbf{I}}

\newcommand{\Co}{\bm{\iota}}

\newcommand{\bP}{\bm{P}}
\newcommand{\bQ}{\bm{Q}}
\newcommand{\bdelta}{\bm{\delta}}
\newcommand{\bphi}{\bm{\phi}}
\newcommand{\bvarphi}{\bm{\varphi}}

\begin{document}

\begin{titlepage}
\renewcommand{\thefootnote}{\fnsymbol{footnote}}

\begin{flushright}
\parbox{3.5cm}
{SNUTP14-013}
\end{flushright}

\vspace*{1.0cm}

\begin{center}
{\LARGE\textbf{Exotic branes and non-geometric fluxes}}%
\end{center}
\vspace{1.0cm}

\centerline{Yuho Sakatani}

\vspace{0.2cm}

\begin{center}
{\it Department of Physics and Astronomy, \\
Seoul National University, Seoul 151-747, KOREA\\}

\vspace{0.2cm}

E-mail: \texttt{yuho@cc.kyoto-su.ac.jp}
\end{center}
\vspace*{1cm}
\begin{abstract}
We present and study ten-dimensional effective actions for various non-geometric fluxes 
of which exotic branes act as the magnetic sources. 
Each theory can be regarded as a $U$-dual version of the $\beta$-supergravity, 
a reformulation of the ten-dimensional supergravity 
which is suitable for describing non-geometric backgrounds with $Q$-flux. 
In each theory, we find a solution that corresponds to the background of an exotic brane 
and show that it is single-valued up to a gauge transformation, 
although the same background written in the standard background fields 
is not single-valued in a usual sense. 
Further, we also find a solution which corresponds to the background 
of an instanton that is the electric dual of the exotic brane and discuss its properties. 
\end{abstract}

\thispagestyle{empty}
\end{titlepage}

\setcounter{footnote}{0}

\section{Introduction}

String theory contains various extended objects such as fundamental strings, 
solitonic five-branes, and $\DD p$-branes. 
These objects are known to couple to the standard background fields; 
the $B$-field or the Ramond-Ramond fields. 
If we consider a compactification on a seven-torus, $T^7_{3\cdots 9}$, 
there arise additional objects, called \emph{exotic branes} 
\cite{Obers:1998fb,Eyras:1999at,LozanoTellechea:2000mc,deBoer:2010ud,Bergshoeff:2011se,deBoer:2012ma}. 
The exotic branes can exist only in the presence of compact isometry directions, 
just like the Kaluza-Klein monopoles, 
and have the tension proportional to $\gs^{\alpha}$ with $\alpha=-2,-3,-4$. 
Among them, a $5^2_2$-brane, which has two isometry directions, has been well-studied recently 
\cite{deBoer:2010ud,Bergshoeff:2011se,Kikuchi:2012za,deBoer:2012ma,Hassler:2013wsa,Geissbuhler:2013uka,Kimura:2013fda,Kimura:2013zva,Chatzistavrakidis:2013jqa,Kimura:2013khz,Andriot:2014uda,Kimura:2014upa,Okada:2014wma}. 
Since the $5^2_2$ background has a non-vanishing (magnetic) \emph{$Q$-flux} \cite{Hassler:2013wsa,Geissbuhler:2013uka}, 
we can identify the $5^2_2$-brane as an object that magnetically couples to a bi-vector field $\beta^{ij}$ 
whose derivative gives the $Q$-flux. 
This can be shown more explicitly by writing down 
the worldvolume effective action of the $5^2_2$-brane \cite{Chatzistavrakidis:2013jqa,Kimura:2014upa}. 

\begin{figure}[b]
\centering
\includegraphics[width=0.78\linewidth]{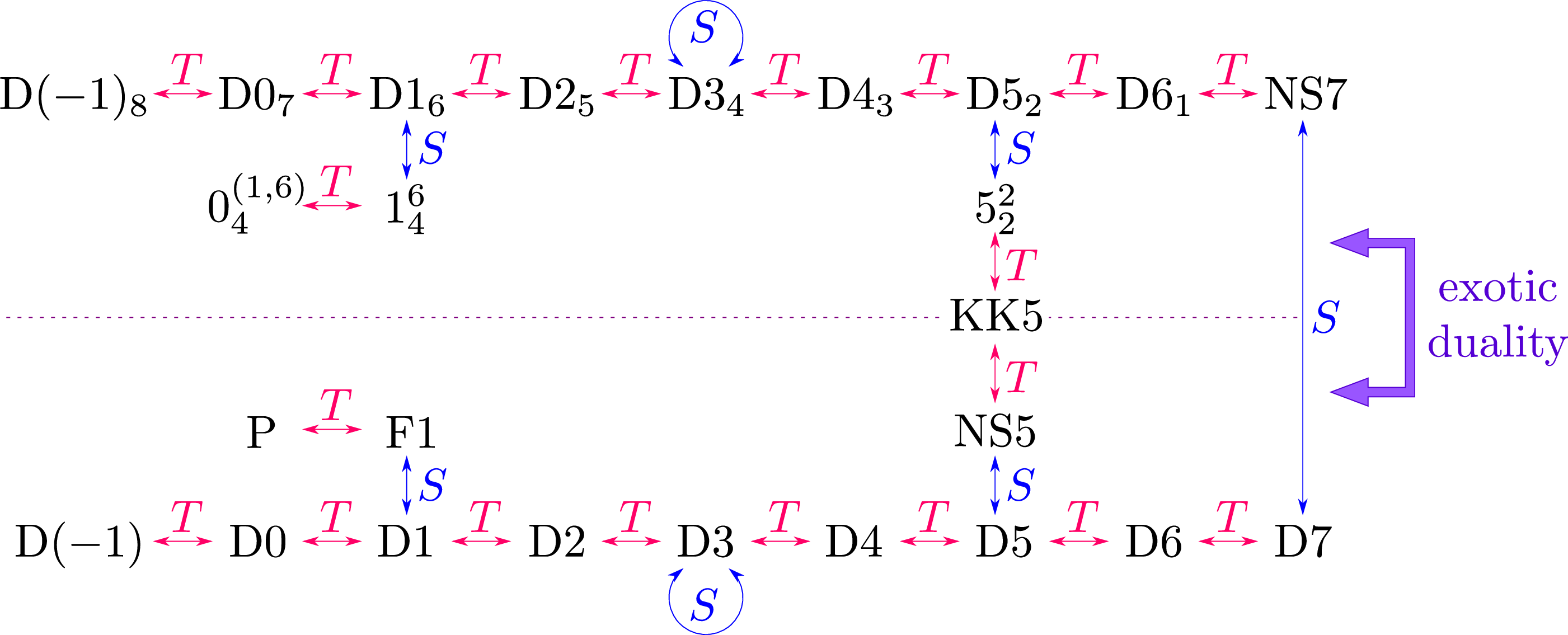}
\caption{A family of exotic branes and the duality web.}
\label{fig:exotic}
\end{figure}

If we perform an $S$-duality transformation, 
the $5^2_2$-brane is mapped to another exotic brane, called a $\DD 5_2$-brane, 
which is a member of a family of exotic $p$-branes, 
denoted by $\DD p_{7-p}$ \cite{Eyras:1999at,LozanoTellechea:2000mc}. 
If we adopt a notation used in \cite{Obers:1998fb,deBoer:2012ma}, 
the exotic $p$-brane, which has $(7-p)$ special isometry directions, 
is denoted by $p_3^{7-p}(n_1\cdots n_p,m_1\cdots m_{7-p})$ 
since the mass is written as 
\begin{align}
 M = \frac{1}{\gs^3\,\ls}\,\Bigl(\frac{R_{n_1}\cdots R_{n_p}}{\ls^p}\Bigr)\,\Bigl(\frac{R_{m_1}\cdots R_{m_{7-p}}}{\ls^{7-p}}\Bigr)^2 \quad
 \bigl(\text{$R_i$: radius in the $x^i$-direction}\bigr)\,,
\end{align}
where $x^{n_i}$ are the extending directions while 
$x^{m_i}$ are the special isometry directions. 
They are also called the higher Kaluza-Klein branes \cite{LozanoTellechea:2000mc}, 
since the quadratic dependence on the radii in the isometry directions 
is similar to the case of the Kaluza-Klein monopole, KK5$=5^1_2$\,. 
For the special case of $p=7$, we frequently denote it by NS7 instead of $7_3$\,. 
The duality relation between the standard branes and the exotic branes 
is summarized in Figure \ref{fig:exotic}. 

In spite of the presence of a symmetric structure between the exotic branes and the usual branes 
(see Figure \ref{fig:exotic}), 
little is known about the exotic branes; 
e.g., the background fields which couple to the exotic branes 
have not been studied in detail, other than the case of the $5^2_2$-brane. 

The main interest in this paper is to identify the background fields 
which couple to the exotic branes 
and to write down the effective supergravity action for the background fields. 
For the $5^2_2$-brane, the relevant background field is a bi-vector $\beta^{ij}$ 
which is a function of the standard NS-NS fields. 
The effective theory for the $\beta$-field has been constructed in a series of works 
\cite{Andriot:2011uh,Andriot:2012wx,Andriot:2012an,Andriot:2013xca,Andriot:2014qla} 
and is called the \emph{$\beta$-supergravity}. 
On the other hand, for the $\DD p_{7-p}$-brane, 
the relevant background field is expected to be a $(7-p)$-vector $\gamma^{i_1\cdots i_{7-p}}$ 
whose derivative is called the non-geometric \emph{$P$-flux} 
(see \cite{Aldazabal:2006up,Aldazabal:2008zza,Aldazabal:2010ef} 
where the $\gamma$-fields are introduced 
in the study of the exceptional generalized geometry, 
\cite{Bergshoeff:2010xc,Bergshoeff:2011zk,Bergshoeff:2011se} where the relation between 
mixed-symmetry tensors and exotic branes is discussed, 
\cite{Chatzistavrakidis:2013jqa,Kimura:2014upa} where the effective $\DD 5_2$-brane action is written down and 
the $\DD 5_2$-brane is found to couple to a bi-vector $\gamma^{ij}$ magnetically, 
and \cite{Andriot:2013xca,Andriot:2014qla} where a possible relation between the polyvectors $\gamma$ and exotic branes is discussed). 
However, the definition of the $\gamma$-fields and the effective action for the $\gamma$-fields are still not fully understood. 

In this paper, assuming the existence of some isometry directions, 
we construct effective actions for various mixed-symmetry tensors that couple to exotic branes. 
We consider the cases of the exotic $5^2_2$-brane, the $1^6_4$-brane, and the $\DD p_{7-p}$-brane, 
and argue that these exotic branes are the magnetic sources of the non-geometric fluxes 
associated with polyvectors 
$\beta^{ij}$, $\beta^{i_1\cdots i_6}$, and $\gamma^{i_1\cdots i_{7-p}}$, respectively. 
As it is well-known, an exotic-brane background written in terms of the usual background fields 
is not single-valued and has a $U$-duality monodromy. 
However, with a suitable redefinition of the background fields, 
the $U$-duality monodromy of the exotic-brane background 
simply becomes a gauge transformation associated with a shift in a polyvector 
(which corresponds to a natural extension of the \emph{$\beta$-transformation} known in the generalized geometry). 
This kind of field redefinition and the rewriting of the action in terms of the new background fields are the main tasks of this paper. 
We further find a new instanton solution that corresponds to the electric source 
of the non-geometric flux, whose existence has been anticipated in section 7 of \cite{Bergshoeff:2011se}. 

This paper is organized as follows. 
In section \ref{sec:supergravity-description}, 
we review the supergravity description of various defect branes 
with an emphasis on the \emph{exotic duality} \cite{Bergshoeff:2011se}, 
which relates the objects described in the upper half (exotic branes) 
and the lower half (standard branes) of Figure \ref{fig:exotic}. 
In section \ref{sec:beta-sugra}, we review the $\beta$-supergravity 
and examine the $5^2_2$-brane as the magnetic source of the $Q$-flux. 
We also find an instanton solution of the $\beta$-supergravity that has the electric charge associated with the $Q$-flux. 
In section \ref{sec:S-dual-bega}, utilizing the techniques of the $\beta$-supergravity, 
we derive the effective action for the bi-vector $\gamma^{ij}$ (whose derivative gives the non-geometric $P$-flux), 
and find a globally well-defined solution that corresponds to the $5^2_3$-brane. 
In section \ref{sec:exotic-p-branes}, 
we derive the effective actions for various polyvectors 
and show that the (simplified) action of the $\beta$-supergravity 
and the action obtained in section \ref{sec:S-dual-bega} are reproduced as special cases. 
We then find two kinds of solutions with either the magnetic or the electric charge associated with the non-geometric flux.
Section \ref{sec:summary} is devoted to summary and discussions. 

\section{Supergravity description of defect branes}
\label{sec:supergravity-description}

In this section, we review the supergravity solutions corresponding to various defect branes, 
and discuss the $\SL(2,\lZ)$ duality \cite{Eyras:1999at,LozanoTellechea:2000mc} 
which relates the standard branes and the exotic branes. 
In the following, we basically follow the notations of \cite{Okada:2014wma}. 

All defect-brane backgrounds considered in this paper can be obtained 
from the following seven-brane background by performing the $T$- and $S$-dualities:
\begin{align}
\begin{split}
 \underline{\text{7-brane:}}\quad & \rmd s^2= \rho_2^{-1/2}\, \bigl(\rho_2\,\abs{f}^2\,\rmd z\,\rmd \bar{z}+\rmd x^2_{03\cdots 9}\bigr) \qquad 
 \bigl(z\equiv r \Exp{\ii\theta}\equiv x^1+\ii x^2\bigr)\,,
\\
 &\Exp{2\phi} = \rho_2^{-2} \,,\quad 
 C^{(0)} = \rho_1 \,, \quad 
 C^{(8)} =-\rho_2^{-1}\,\rmd t\wedge \rmd x^3\wedge \cdots \wedge \rmd x^9\,.
\end{split}
\label{eq:7-branes}
\end{align}
This background satisfies the equations of motion for the type IIB supergravity 
as long as the functions $\rho \equiv \rho_1+\ii\rho_2$ 
and $f$ are holomorphic functions of $z$. 
In the following, we choose them as $\rho(z)=\ii (\sigma/2\pi)\,\log(r_\text{c}/z)= (\sigma/2\pi)\,\bigl[\theta+\ii \log(r_\text{c}/r)\bigr]$ 
($r_\text{c}\,$: positive constant) and $f(z)=1$, 
which makes the above background the well-known D7 background. 
Here, the value of $\sigma$ depends on the duality frame (see Table \ref{tab:sigma}) 
and it is now given by the string coupling constant; $\sigma_{\DD7}=\gs$\,. 
\begin{table}[b]
\centering
 \begin{tabular}{|c||l||c||l|}\hline
 $\DD p$ & $\sigma_{\DD p}=\gs\,\bigl(\ls^{7-p}/R_{m_1} \cdots R_{m_{7-p}}\bigr)$ & 
 $p^{7-p}_3$ & $\sigma_{p^{7-p}_3}=\gs^{-1}\,\bigl(R_{m_1} \cdots R_{m_{7-p}}/\ls^{7-p}\bigr)$ 
\\\hline
 $\NS5$ & $\sigma_{\NS5}= \ls^2/R_{m_1}R_{m_2}$ & 
 $5^2_2$ & $\sigma_{5^2_2}= R_{m_1}R_{m_2}/\ls^2$ 
\\\hline
 $\KK 5$ & $\sigma_{\KK 5}= R_m/R_\ell$ & \multicolumn{2}{c|}{---}
\\\hline
 F1 & $\sigma_{\FF1}= \gs^2\,\bigl(\ls^6/R_{m_1}\cdots R_{m_6}\bigr)$ &
 $1^6_4$ & $\sigma_{1^6_4}= \gs^{-2}\,\bigl(R_{m_1}\cdots R_{m_6}/\ls^6\bigr)$ 
\\\hline
 P & $\sigma_{\PP}=\gs^2\,\bigl(\ls^8/R_{m_1}\cdots R_{m_6}R_n^2\bigr)$ &
 $0^{(1,6)}_4$ & $\sigma_{0^{(1,6)}_4}=\gs^{-2}\,\bigl(R_{m_1}\cdots R_{m_6}R_n^2/\ls^8\bigr)$ 
\\\hline 
 \end{tabular}
\caption{Values of the dimensionless constant $\sigma$ in various duality frames. 
The directions $x^n$, $x^{m_i}$, and $x^\ell$ are the same as those appearing in the 
expressions for the background fields.}
\label{tab:sigma}
\end{table}
Note that in the solutions described below, 
we can always redefine the holomorphic functions as (see \cite{Bergshoeff:2006jj})
\begin{align}
 \rho(z) \to \rho'(z) = \frac{a\,\rho(z) + b}{c\,\rho(z) + d}\,,\quad 
 f(z)\to f'(z)=(c\,\rho + d)\,f(z)\,,\quad 
 \bpm a & b \\ c & d \epm \in \SL(2,\lZ)\,,
\label{eq:SL2Z}
\end{align}
which corresponds to the $\SL(2,\lZ)$ symmetry in the type IIB theory. 
Note also that all backgrounds considered in this paper, by construction, 
have isometries in the $03\cdots 9$-directions. 

The defect D$p(n_1\cdots n_p)$ background, 
which corresponds to a D$p$-brane extending in the $x^{n_1},\dotsc,x^{n_p}$-directions 
and smeared over the remaining directions, $x^{m_1},\dotsc,x^{m_{7-p}}$, of the seven-torus $T^7_{3\cdots 9}$ ($n_i,m_i=3,\dotsc,9$), 
is given by
\begin{align}
\begin{split}
 \underline{\text{ $\DD p$ :}}\quad &\rmd s^2
  = \rho_2^{-1/2}\, \bigl(\rho_2\,\abs{f}^2\,\rmd z\,\rmd \bar{z}+\rmd x^2_{0n_1\cdots n_p}\bigr) + \rho_2^{1/2}\, \rmd x^2_{m_1\cdots m_{7-p}} \,, 
\\
 &C^{(7-p)} = \rho_1\,\rmd x^{m_1}\wedge\cdots\wedge\rmd x^{m_{7-p}}\,,\quad 
 \Exp{2\phi} = \rho_2^{\frac{3-p}{2}} \,,
\\
 &C^{(p+1)} = - \rho_2^{-1}\,\epsilon_{0n_1\cdots n_pm_{7-p}\cdots m_1}\,\rmd t \wedge \rmd x^{n_1}\wedge \cdots \wedge \rmd x^{n_p} \,,
\end{split}
\label{eq:Dp-brane}
\end{align}
where the totally antisymmetric symbol is given by $\epsilon_{03\cdots 9}\equiv 1$ and the indices are not summed. 
On the other hand, the exotic $p$-brane background, 
$\DD p_{7-p}(n_1\cdots n_p,m_1\cdots m_{7-p})$, is given by
\begin{align}
\begin{split}
 \underline{\text{ $\DD p_{7-p}$ :}}\quad 
 &\rmd s^2 = \frac{\abs{\rho}}{\rho_2^{1/2}}\, \bigl(\rho_2\,\abs{f}^2\,\rmd z\,\rmd \bar{z}+\rmd x^2_{0n_1\cdots n_p}\bigr) + \frac{\rho_2^{1/2}}{\abs{\rho}}\, \rmd x^2_{m_1\cdots m_{7-p}} \,, 
\\
 &C^{(7-p)} = -\frac{\rho_1}{\abs{\rho}^2}\,\rmd x^{m_1}\wedge \cdots \wedge \rmd x^{m_{7-p}}\,, \quad 
 \Exp{2\phi} = \biggl(\frac{\rho_2}{\abs{\rho}^2}\biggr)^{\frac{3-p}{2}} \,,
\\
 &C^{(p+1)} = -\frac{\abs{\rho}^2}{\rho_2}\,\epsilon_{0n_1\cdots n_pm_{7-p}\cdots m_1}\,\rmd t \wedge \rmd x^{n_1}\wedge \cdots \wedge \rmd x^{n_p} \,.
\label{eq:KKp-brane}
\end{split}
\end{align}
As it has been noticed in \cite{Eyras:1999at,LozanoTellechea:2000mc}, this background is obtained from the D$p$-brane background 
through the replacement
\begin{align}
 \rho(z) \to -\rho^{-1}(z) \,,\quad \rho_2\,\abs{f}^2\to \rho_2\,\abs{f}^2\,,
\label{eq:exotic-dual}
\end{align}
which is a special case of the $\SL(2,\lZ)$ transformation given in \eqref{eq:SL2Z}. 
This kind of duality between a usual brane and an exotic brane is called the exotic duality \cite{Bergshoeff:2011se}.

The defect (or smeared) $\NS5(n_1\cdots n_5)$ and the $5^2_2(n_1\cdots n_5,m_1m_2)$ backgrounds are also related 
to each other through the exotic-duality transformation \eqref{eq:exotic-dual}:
\begin{align}
\begin{split}
 \underline{\text{ $\NS5$ :}}\quad 
 &\rmd s^2 = \rho_2\,\abs{f}^2\,\rmd z\,\rmd \bar{z} + \rmd x^2_{0n_1\cdots n_5} + \rho_2\, \rmd x_{m_1m_2}^2 \,, \quad
 \Exp{2\phi} = \rho_2 \,, 
\\
 &B^{(2)}= \rho_1\,\rmd x^{m_1}\wedge \rmd x^{m_2}\,, \quad 
 B^{(6)}= \rho_2^{-1}\,\epsilon_{0n_1\cdots n_5m_2m_1}\,\rmd t\wedge \rmd x^{n_1}\wedge\cdots\wedge \rmd x^{n_5}\,,
\label{eq:NS5}
\end{split}
\\
\begin{split}
 \underline{\text{ $5^2_2$ :}}\quad 
 &\rmd s^2 = \rho_2\,\abs{f}^2\,\rmd z\,\rmd \bar{z} +\rmd x^2_{0n_1\cdots n_5} + \frac{\rho_2}{\abs{\rho}^2}\,\, \rmd x_{m_1m_2}^2 \,, \quad
 \Exp{2\phi} = \frac{\rho_2}{\abs{\rho}^2}\,,
\\
 &B^{(2)} = -\frac{\rho_1}{\abs{\rho}^2}\,\rmd x^{m_1}\wedge \rmd x^{m_2} \,, \quad 
 B^{(6)}= \frac{\abs{\rho}^2}{\rho_2}\,\epsilon_{0n_1\cdots n_5m_2m_1}\,\rmd t\wedge \rmd x^{n_1}\wedge\cdots\wedge \rmd x^{n_5}\,.
\label{eq:522}
\end{split}
\end{align}
Applying a general $\SL(2,\lZ)$ transformation to these five-brane backgrounds, 
we can obtain the background of a \emph{defect $(p,q)$-five brane} 
\cite{Kimura:2014wga,deBoer:2012ma}, 
which is a bound state of $p$ defect NS5-branes and $q$ $5^2_2$-branes. 
Note that we can perform the $\SL(2,\lZ)$ transformation even in the type IIA theory, 
since, in this duality frame, the $\SL(2,\lZ)$ transformation 
is realized as a subgroup of the $T$-duality group. 

The background of a defect KK5$(n_1\cdots n_5,m)$-brane smeared in the $x^\ell$-direction 
and its exotic-dual background are given by
\begin{align}
\begin{split}
 \underline{\text{ $\KK 5$ :}}\quad 
 &\rmd s^2 = \rho_2\,\abs{f}^2\,\rmd z\,\rmd \bar{z} + \rmd x^2_{0n_1\cdots n_5} +\rho_2\,\rmd x_\ell^2 + \rho_2^{-1}\,\bigl(\rmd x^m -\rho_1\,\rmd x^\ell\bigr)^2  \,,
\\
 &\Exp{2\phi} =1\,,\quad B^{(2)} = 0 \,,
\end{split}
\\
\begin{split}
 \underline{\text{ anti-$\KK5$ :}}\quad 
 &\rmd s^2 = \rho_2\,\abs{f}^2\,\rmd z\,\rmd \bar{z} + \rmd x^2_{0n_1\cdots n_5} +\rho_2\,\rmd x_m^2 + \rho_2^{-1}\,\bigl(\rmd x^\ell +\rho_1\,\rmd x^m\bigr)^2 \,,
\\
 &\Exp{2\phi} =1\,,\quad B^{(2)} = 0 \,. 
\end{split}
\end{align}
The latter corresponds to the background of an anti-KK5$(n_1\cdots n_5,\ell)$-brane smeared in the $x^m$-direction. 
Namely, under the exotic-duality transformation, the Taub-NUT direction is interchanged with the smeared direction. 
The background of a bound state of these KK5-branes is also considered in \cite{Kimura:2014wga}. 

Further, there are the following pairs of strings and pp-waves:
\begin{align}
\begin{split}
 \underline{\text{ \FF1 :}}\quad 
 & \rmd s^2 = \rho_2^{-1}\, \bigl(\rho_2\,\abs{f}^2\,\rmd z\,\rmd \bar{z} +\rmd x_{0n}^2 \bigr)+ \rmd x^2_{m_1\cdots m_6} \,, \quad 
    \Exp{2\phi} = \rho_2^{-1} \,,
\\
   &B^{(6)} = \rho_1\,\rmd x^{m_1}\wedge \cdots \wedge \rmd x^{m_{6}}\,, \quad 
    B^{(2)} = -\rho_2^{-1}\,\epsilon_{0nm_1\cdots m_6}\,\rmd t \wedge \rmd x^n \,,
\end{split}
\label{eq:F1-BG}
\\
\begin{split}
 \underline{\text{ $1^6_4$ :}}\quad &\rmd s^2= \frac{\abs{\rho}^2}{\rho_2} \, \bigl(\rho_2\,\abs{f}^2\,\rmd z\,\rmd \bar{z} +\rmd x_{0n}^2 \bigr)+ \rmd x^2_{m_1\cdots m_6} \,, \quad 
 \Exp{2\phi} = \frac{\abs{\rho}^2}{\rho_2} \,,
\\
   &B^{(6)}= -\frac{\rho_1}{\abs{\rho}^2}\,\rmd x^{m_1}\wedge \cdots \wedge \rmd x^{m_{6}}\,, \quad 
    B^{(2)} = -\frac{\abs{\rho}^2}{\rho_2}\,\epsilon_{0nm_1\cdots m_6}\,\rmd t \wedge \rmd x^n \,,
\label{eq:1^6_4-BG}
\end{split}
\\
\begin{split}
 \underline{\text{ P :}}\quad & \rmd s^2 = -2\,\rmd t\,\rmd x^n + \rho_2\, \rmd x_n^2 + \abs{f}^2\,\rmd z\,\rmd \bar{z} + \rmd x^2_{m_1\cdots m_6} \,, 
\\
 &\Exp{2\phi} = \rho_2^{-1} \,,\quad B^{(2)} = 0\,, 
\end{split}
\\
\begin{split}
 \underline{\text{ $0^{(1,6)}_4$ :}}\quad 
 & \rmd s^2 = -2\,\rmd t\,\rmd x^n + \frac{\rho_2}{\abs{\rho}^2}\, \rmd x_n^2 + \abs{\rho}^2\,\abs{f}^2\,\rmd z\,\rmd \bar{z} + \rmd x^2_{m_1\cdots m_6} \,, 
\\
 &\Exp{2\phi} = \frac{\abs{\rho}^2}{\rho_2} \,,\quad 
 B^{(2)} = 0\,. 
\end{split}
\end{align}

If we perform a timelike $T$-duality in the $\DD 0$ or the $\DD 0_7$ background, 
we obtain the following defect D-instanton (or more precisely the defect E0-brane \cite{Hull:1998vg}) background or another instanton background, 
to be called $\DD(-1)_8$:
\begin{align}
\begin{split}
 \underline{\text{ $\DD(-1)$ \ [IIB$^\star$]\,:}}\quad 
 & \rmd s^2 = \Exp{\phi/2}\, \bigl(\abs{f}^2\,\rmd z\,\rmd \bar{z}+ \rmd x^2_{03\cdots 9}\bigr) \,, \quad 
   \Exp{2\phi} = \rho_2^2 \,, 
\\
 &C^{(8)} = \rho_1\,\rmd t\wedge\rmd x^3\wedge\cdots\wedge\rmd x^9\,,\quad C^{(0)} = \rho_2^{-1} \,, 
\end{split}
\\
\begin{split}
 \underline{\text{ $\DD(-1)_8$ \ [IIB$^\star$]\,:}}\quad 
 & \rmd s^2 = \Exp{\phi/2}\, \bigl(\abs{\rho}^2\,\abs{f}^2\,\rmd z\,\rmd \bar{z}+ \rmd x^2_{03\cdots 9}\bigr) \,, \quad 
  \Exp{2\phi} = \frac{\rho_2^2}{\abs{\rho}^4} \,,
\\
 &C^{(8)} = -\frac{\rho_1}{\abs{\rho}^2}\,\rmd t\wedge\rmd x^3\wedge\cdots\wedge\rmd x^9\,,\quad C^{(0)} = \frac{\abs{\rho}^2}{\rho_2} \,.
\end{split}
\end{align}
These backgrounds satisfy the equations of motion for the type IIB$^\star$ theory of \cite{Hull:1998vg} 
since we have performed a timelike $T$-duality. 
The corresponding backgrounds written as solutions of the (\emph{Euclideanised}) type IIB theory 
are given by (note that the Euclideanisation is given by the replacements 
$t\to \ii\tau$ and $C^{(0)}\to \ii C^{(0)}$, and then $C^{(8)}$ is defined by $\rmd C^{(8)}=-*\rmd C^{(0)}$)
\begin{align}
\begin{split}
 \underline{\text{ $\DD(-1)$ \ [IIB]\,:}}\quad 
 & \rmd s^2 = \Exp{\phi/2}\, \bigl(\abs{f}^2\,\rmd z\,\rmd \bar{z}+\rmd\tau^2+\rmd x^2_{3\cdots 9}\bigr) \,, \quad 
   \Exp{2\phi} = \rho_2^2 \,, 
\\
 &C^{(8)} = -\rho_1\,\rmd \tau\wedge\rmd x^3\wedge\cdots\wedge\rmd x^9\,,\quad 
 C^{(0)} = \rho_2^{-1} \,, 
\label{eq:D-inst-IIB}
\end{split}
\\
\begin{split}
 \underline{\text{ $\DD(-1)_8$ \ [IIB]\,:}}\quad 
 & \rmd s^2 = \Exp{\phi/2}\, \bigl(\abs{\rho}^2\,\abs{f}^2\,\rmd z\,\rmd \bar{z}+\rmd\tau^2+ \rmd x^2_{3\cdots 9}\bigr) \,, \quad 
   \Exp{2\phi} = \frac{\rho_2^2}{\abs{\rho}^4} \,,
\\
 &C^{(8)} = \frac{\rho_1}{\abs{\rho}^2}\,\rmd \tau\wedge\rmd x^3\wedge\cdots\wedge\rmd x^9\,,\quad 
 C^{(0)} = \frac{\abs{\rho}^2}{\rho_2} \,.
\end{split}
\label{eq:D-inst-8}
\end{align}
Note that these objects are not related to each other by an $S$-duality, 
as opposed to the case of seven branes.  
Further details about instanton backgrounds are discussed in section \ref{sec:p=7}. 

The uplifts of the above defect backgrounds to eleven dimensions are given in 
\cite{LozanoTellechea:2000mc,Bergshoeff:2011se,deBoer:2012ma} 
although we do not consider them in this paper. 

\section{$\beta$-supergravity}
\label{sec:beta-sugra}

Recently, it has been pointed out in \cite{deBoer:2010ud} 
that the $5^2_2$ background \eqref{eq:522} has a $T$-duality monodromy around the center. 
That is, globally, we have to glue the background fields in different coordinate patches 
using a $T$-duality transformation. 
This kind of \emph{non-geometric background} which requires to use a $T$-duality transformation 
as a transition function is called a \emph{$T$-fold}. 
More generally, a non-geometric background which requires to use 
a larger duality transformation in string theory, i.e.~the $U$-duality symmetry, 
is called a \emph{$U$-fold} \cite{Hull:2004in}. 

In the special case of $T$-folds, 
we can globally describe the backgrounds within a framework of the double field theory (DFT), 
which is a $T$-duality covariant reformulation of the low-energy supergravity theory 
\cite{Hull:2009mi,Hull:2009zb,Hohm:2010jy,Hohm:2010pp,Hohm:2011dv,Hohm:2010xe,Jeon:2010rw,Jeon:2011cn,Jeon:2011vx,Hohm:2011nu,Jeon:2011sq,Jeon:2012kd,Hohm:2012gk,Jeon:2012hp,Berman:2013uda}. 
In DFT, in addition to the usual spacetime coordinates $x^i$ ($i =0,1,\dotsc,9$), 
we also introduce the ``dual'' coordinates $\tilde{x}_i$, 
and treat them on an equal footing; $(x^I)\equiv (\tilde{x}_i,\,x^i)$\,.
The fundamental fields of DFT are the generalized metric 
\begin{align}
 (\cH_{IJ}) \equiv 
 \begin{pmatrix} G^{-1} & -G^{-1}\,B \\ B\, G^{-1} & G-B\, G^{-1}\,B \end{pmatrix} \quad 
 \bigl[G\equiv (G_{ij})\,,\quad B\equiv(B_{ij})\bigr]\,,
\end{align}
and the ($T$-duality invariant) dilaton $\Exp{-2d}\equiv \sqrt{\abs{G}}\,\Exp{-2\phi}$. 
According to a constraint, called the strong constraint, 
these fields can depend only on the half of the coordinates. 
If we choose the background fields to depend only on the usual coordinates $x^i$, 
the DFT action reduces to the standard ten-dimensional action for the NS-NS fields. 
In DFT, a choice of coordinates corresponds to a choice of a $T$-duality frame, 
and indeed, a \emph{generalized coordinate transformation}, 
which is a gauge symmetry of DFT, corresponds to an $\OO(10,10)$ transformation. 
More explicitly, under a coordinate transformation, $x^I\to x^{\prime\,I}(x^I)$, 
the generalized metric (which is a \emph{generalized tensor}) transforms as \cite{Hohm:2012gk}
\begin{align}
 \cH'_{IJ} = \mathcal{F}_I{}^K\,\mathcal{F}_J{}^L\,\cH_{KL} \,,\quad 
 \cF_I{}^J 
 &\equiv \frac{1}{2}\,\Bigl(\frac{\partial x^K}{\partial x^{\prime\,I}}\frac{\partial x'_K}{\partial x_J}
        +\frac{\partial x'_I}{\partial x_K}\frac{\partial x^J}{\partial x^{\prime\,K}}\Bigr) \,,
\end{align}
where indices are raised or lowered by using the $\OO(10,10)$-invariant metric, 
$\eta = \bigl(\begin{smallmatrix} 0& \mathbf{1} \cr \mathbf{1} &0\end{smallmatrix}\bigr)$\,.

Now, let us consider a generalized coordinate transformation 
\begin{align}
 x^{\prime\,i} = \tilde{x}_i\,,\quad \tilde{x}'_i =x^i \,,
\label{eq:dual-map}
\end{align}
in a standard background that does not depend on the dual coordinates $\tilde{x}_i$\,. 
Since $\cF_I{}^J=\delta_I^J$, the functional form of the generalized metric $\cH_{IJ}$ is invariant, 
although the background fields become functions only of the dual coordinates $\tilde{x}_i$\,. 
In the ``dual'' spacetime, spanned by the dual coordinates $\tilde{x}_i$\,, 
a natural set of background fields,%%%%%%%%%%%%%%%%%%%%%%%%%%%%%%%%%%%%%%%%%%%%%%%%%%%%%%%
\footnote{For example, the string sigma model action for the dual coordinates $\tilde{X}_i(\sigma)$ can be written as (see e.g.~\cite{Hull:2004in})
\begin{align*}
 S=- \frac{1}{4\pi\ap}\,\int\rmd^2\sigma\sqrt{-\eta}\,\bigl(\eta^{\alpha\beta}\,\tilde{G}^{ij} 
   + \epsilon^{\alpha\beta}\,\tilde{B}^{ij}\bigr) \,\partial_\alpha \tilde{X}_i\, \partial_\beta \tilde{X}_j\,. 
\end{align*}} 
%%%%%%%%%%%%%%%%%%%%%%%%%%%%%%%%%%%%%%%%%%%%%%%%%%%%%%%%%%%%%%%%%%%%%%%%%%%%%%%%%%%%%%%%%%
which we denote by $(\tilde{G}^{ij},\,\tilde{\varphi},\,\tilde{B}^{ij})$, is given by
\begin{align}
\begin{split}
 \tilde{G}^{ij}&\equiv \bigl(E^{-1}\bigr)^{ik}\,\bigl(E^{-1}\bigr)^{jl}\,G_{kl} \,, \quad
 E\equiv \bigl(E_{ij}\bigr)\equiv \bigl(G_{ij} + B_{ij}\bigr) \,,
\\
 \tilde{B}^{ij}&\equiv \bigl(E^{-1}\bigr)^{ik}\,\bigl(E^{-1}\bigr)^{jl}\,B_{kl}\,, \quad 
 \Exp{-2\tilde{\varphi}}\sqrt{\abs{\tilde{G}}}\equiv \Exp{-2\phi}\sqrt{\abs{G}} \,.
\end{split}
\label{eq:tilde-T-dual}
\end{align}
These ``dual'' background fields satisfy the standard ten-dimensional equations of motion for the NS-NS fields
with the following replacements:
\begin{align}
 x^i \to \tilde{x}_i\,,\quad \partial_i\to \tilde{\partial}^i\,,\quad 
 G_{ij}\to \tilde{G}^{ij} \,,\quad B_{ij}\to \tilde{B}^{ij}\,, \quad 
 \sqrt{\abs{G}}\Exp{-2\phi} \to \sqrt{\abs{\tilde{G}}}\Exp{-2\tilde{\varphi}} \,.
\end{align}

Under the generalized coordinate transformation \eqref{eq:dual-map}, the $5^2_2(34567,89)$ background 
is mapped to the dual background;
\begin{align}
\begin{split}
 \underline{\text{ $5^2_2$ :}}\quad 
 \rmd \tilde{s}^2 &\equiv \tilde{G}^{ij}\,\rmd \tilde{x}_i\,\rmd \tilde{x}_j
 = \rho_2^{-1}\,\abs{f}^{-2}\,\rmd z\,\rmd\bar{z} + \rmd \tilde{x}^2_{034567} + \rho_2\, \rmd \tilde{x}_{89}^2 \,, 
\\
 \tilde{B}^{(2)} &\equiv \frac{1}{2}\,\tilde{B}^{ij}\,\rmd \tilde{x}_i\wedge \rmd \tilde{x}_j
 = -\rho_1\,\rmd \tilde{x}_8\wedge \rmd \tilde{x}_9 \,,\quad 
 \Exp{2\tilde{\varphi}} = \abs{f}^{-2}\,\rho_2^{-1} \,.
\label{eq:522-dual}
\end{split}
\end{align}
This background has the same form with the (geometric) $\NS5(34567)$ background 
in the usual ten-dimensional spacetime 
(apart from the $1$-$2$ components of the metric and the dilaton). 
That is, the $5^2_2$ background written in the dual fields \eqref{eq:522-dual} 
is a geometric background in a sense that 
the monodromy, $\rho_1\to \rho_1+\sigma$, is just a gauge transformation associated with the $\tilde{B}$-field.%%%%%%
\footnote{%%%%%%%%%%%%%%%%%%%%%%%%%%%%%%%%%%%%%%%%%%%%%%%%%%%%%%%%%%%%%%%%%%%%%%%%%%%%%%%
See section 5.1 of \cite{Okada:2014wma} for another explanation that 
the $5^2_2$ background (described as a doubled geometry) is geometric; 
the monodromy is realized as a generalized diffeomorphism there.} 
%%%%%%%%%%%%%%%%%%%%%%%%%%%%%%%%%%%%%%%%%%%%%%%%%%%%%%%%%%%%%%%%%%%%%%%%%%%%%%%%%%%%%%%%%
The redefinition of the background fields \eqref{eq:tilde-T-dual} is essential 
in the formulation of the $\beta$-supergravity, 
although we do not introduce the dual coordinates $\tilde{x}_i$ in the $\beta$-supergravity. 

\subsection{The action and its reduction}

In the $\beta$-supergravity, the fundamental fields 
$(\tilde{g}_{ij},\,\tilde{\phi},\,\tilde{\beta}^{ij})$ (which depend only on the usual ten-dimensional coordinates $x^i$) 
are defined by
\begin{align}
 \tilde{g}_{ij}&\equiv E_{ik}\,E_{jl}\,G^{kl} \,, \quad
 \tilde{\beta}^{ij} \equiv \bigl(E^{-1}\bigr)^{ik}\,\bigl(E^{-1}\bigr)^{jl}\,B_{kl}\,,\quad 
 \Exp{-2\tilde{\phi}}\sqrt{\abs{\tilde{g}}}\equiv \Exp{-2\phi}\sqrt{\abs{G}}\,.
\label{eq:beta-fields}
\end{align}
Note that the metric $\tilde{g}_{ij}$ is the inverse of 
the dual metric $\tilde{G}^{ij}$ introduced in \eqref{eq:tilde-T-dual}, 
and accordingly the dilaton $\tilde{\phi}$ is different from 
$\tilde{\varphi}$ given in \eqref{eq:tilde-T-dual}. 
Further, note that $\tilde{\beta}^{ij}$ behaves as a bi-vector under diffeomorphisms 
(in the standard ten-dimensional spacetime) 
although its functional form is the same with the ``2-form'' $\tilde{B}^{ij}$ in the dual spacetime. 
In a series of papers \cite{Andriot:2011uh,Andriot:2012wx,Andriot:2012an,Andriot:2013xca}, 
an effective action for $(\tilde{g}_{ij},\,\tilde{\phi},\,\tilde{\beta}^{ij})$ was proposed 
and it was shown that the action is equal to the usual action for the NS-NS fields (up to boundary terms) 
if we rewrite these fields using the original background fields through the relation \eqref{eq:beta-fields}. 
The effective action, written in a manifestly invariant form under diffeomorphisms, is given by \cite{Andriot:2013xca}
\begin{align}
\begin{split}
 S\bigl[\,\tilde{g}_{ij},\,\tilde{\phi},\,\tilde{\beta}^{ij}\,\bigr]
 &= \frac{1}{2\kappa_{10}^2}\,\int\rmd^{10}x\sqrt{\abs{\tilde{g}}}\,\Exp{-2\tilde{\phi}}\,
   \Bigl(\tilde{R}+4\,\tilde{g}^{ij}\,\partial_i \tilde{\phi}\,\partial_j \tilde{\phi} + \widecheck{\cR}(\tilde{g}_{ij},\,\tilde{\beta}^{ij}) 
\\
 &\qquad\qquad\qquad\qquad\qquad\qquad - \frac{1}{2}\,\abs{R^{ijk}}^2 
    + 4\,\abs{\tilde{\beta}^{ij}\,\partial_j \tilde{\phi} -\cT^i(\tilde{g}_{ij},\,\tilde{\beta}^{ij})}^2\Bigr)\,,
\label{eq:beta-action}
\end{split}
\end{align}
where we defined
\begin{align}
\begin{split}
 &\widecheck{\cR}^{ij}(\tilde{g}_{ij},\,\tilde{\beta}^{ij})\equiv 
 -\tilde{\beta}^{kl}\,\partial_l\widecheck{\Gamma}^{ij}_k+\tilde{\beta}^{il}\,\partial_l\widecheck{\Gamma}^{kj}_k
 +\widecheck{\Gamma}^{ij}_k\,\widecheck{\Gamma}^{lk}_l-\widecheck{\Gamma}^{li}_k\,\widecheck{\Gamma}^{kj}_l\,,
\\
 &\widecheck{\cR}(\tilde{g}_{ij},\,\tilde{\beta}^{ij})\equiv \tilde{g}_{ij}\,\widecheck{\cR}^{ij}(\tilde{g}_{ij},\,\tilde{\beta}^{ij})\,, 
\\
 &\widecheck{\Gamma}^{ij}_k \equiv \frac{1}{2}\,\tilde{g}_{kl}\,\bigl(-\tilde{\beta}^{im}\,\partial_m\tilde{g}^{jl}-\tilde{\beta}^{jm}\,\partial_m\tilde{g}^{il}+\tilde{\beta}^{lm}\,\partial_m\tilde{g}^{ij}\bigr) +\tilde{g}_{kl}\,\tilde{g}^{m(i}\,\partial_m\tilde{\beta}^{j)l} -\frac{1}{2}\,\partial_k\tilde{\beta}^{ij} \,,
\\
 &\cT^i(\tilde{g}_{ij},\,\tilde{\beta}^{ij}) \equiv \widecheck{\Gamma}^{ki}_k = \partial_k\tilde{\beta}^{ik}-\frac{1}{2}\,\tilde{\beta}^{ij}\,\tilde{g}_{kl}\,\partial_j\tilde{g}^{kl}
 = \nabla_k\tilde{\beta}^{ik}\,,
\\
 &R^{ijk}\equiv 3\,\tilde{\beta}^{l[i}\partial_l\tilde{\beta}^{jk]}\,, \quad 
  \abs{V^{i_1\cdots i_p}}^2\equiv \frac{1}{p!}\,\tilde{g}_{i_1j_1}\cdots \tilde{g}_{i_pj_p}\, V^{i_1\cdots i_p}\,V^{j_1\cdots j_p}\,.
\end{split}
\label{eq:beta-definitions}
\end{align}

Note that we can assume that any derivative $\partial_i$ contracted with $\tilde{\beta}^{ij}$ vanishes, 
as long as we consider the defect backgrounds which satisfy $\tilde{\beta}^{\bfa i}=0$ ($\bfa=1,2$) and
have isometries in the $03\cdots 9$-directions. 
Under the assumption, the above action reduces to the following simple form \cite{Andriot:2011uh}:
\begin{align}
 S\bigl[\,\tilde{g}_{ij},\,\tilde{\phi},\,\tilde{\beta}^{ij}\,\bigr] 
 = \frac{1}{2\kappa_{10}^2}\, \int \Exp{-2\tilde{\phi}}\,\Bigl(\tilde{*}\,\tilde{R} + 4\,\rmd\tilde{\phi}\wedge \tilde{*}\,\rmd\tilde{\phi} 
  -\frac{1}{4}\,\tilde{g}_{ik}\,\tilde{g}_{jl}\,\bQ^{(1)\,ij}\wedge \tilde{*}\,\bQ^{(1)\,kl} \Bigr) \,,
\label{eq:beta-action-reduced}
\end{align}
where the $Q$-flux, $\bQ^{(1)\,ij}\equiv Q_k{}^{ij}\,\rmd x^k \equiv \rmd \tilde{\beta}^{ij}$, 
is a mixed-symmetry tensor (i.e.~a bi-vector-valued 1-form).%%%%%%%%%%%%%%%%%%%%%%%%%%%
\footnote{In fact, this is not a tensor as we can see from the definition. 
However, if we assume $\tilde{\beta}^{\bfa i}=0$ ($\bfa=1,2$) and the existence 
of isometries in the $03\cdots 9$-directions, $Q_k{}^{ij}$ transforms as a tensor 
under diffeomorphisms, $x'^\bfa=x'^\bfa(x^1,x^2)$ and $x'^\bfp=\Lambda^\bfp{}_\bfq\,x^\bfq$ ($\bfp,\bfq=0,3,\dotsc,9$,\ $\Lambda^\bfp{}_\bfq\in \GL(8,\lR)$)\,, 
which respect the assumptions.
\label{foot:tensor}} 
%%%%%%%%%%%%%%%%%%%%%%%%%%%%%%%%%%%%%%%%%%%%%%%%%%%%%%%%%%%%%%%%%%%%%%%%%%%%%%%%%%%%%%%
In this case, the Bianchi identity and the equation of motion for $\tilde{\beta}^{ij}$ become
\begin{align}
 \rmd \bQ^{(1)\,ij} = 0\,,\quad \rmd \bQ^{(9)}_{ij} =0 \,,\quad 
 \bQ^{(9)}_{ij} \equiv \Exp{-2\tilde{\phi}}\,\tilde{g}_{ik}\,\tilde{g}_{jl}\,\tilde{*}\,\bQ^{(1)\,kl} \,. 
\label{eq:beta-Bianchi}
\end{align}

Assuming the isometries in the $03\cdots9$-directions 
and that the background fields satisfy 
$\tilde{g}_{\bfa\bfp}=0$ and $\tilde{\beta}^{\bfa i}=0$ ($\bfa=1,2$, $\bfp=0,3,\dotsc,9$), 
we can show that $\bQ^{(9)}_{ij}$ has the following form:
\begin{align}
 \bQ^{(9)}_{ij} = \bigl[\bm{q}^{1}_{ij}(x^1,x^2)\,\rmd x^1+\bm{q}^2_{ij}(x^1,x^2)\,\rmd x^2\bigr]\wedge \rmd t\wedge \rmd x^3\wedge\cdots\wedge\rmd x^9\,. 
\end{align}
Then, using $\rmd \bQ^{(9)}_{ij} =0$, we can always choose a gauge 
in which the dual potential $\tilde{\beta}^{(8)}_{ij}$, 
defined by $\bQ^{(9)}_{ij}\equiv \rmd \tilde{\beta}^{(8)}_{ij}$, is proportional to $\rmd t\wedge \rmd x^3\wedge\cdots\wedge\rmd x^9$\,;
\begin{align}
\begin{split}
 \tilde{\beta}^{(8)}_{ij} &= \bm{q}^0_{ij}(x^1,x^2)\,\rmd t\wedge \rmd x^3\wedge\cdots\wedge\rmd x^9 
\\
 \bigl[\rmd \bm{q}^0_{ij}&\equiv \bm{q}^1_{ij}(x^1,x^2)\,\rmd x^1+\bm{q}^2_{ij}(x^1,x^2)\,\rmd x^2\bigr]\,.
\label{eq:beta-8}
\end{split}
\end{align}
If we define a contraction of a mixed-symmetry tensor $\tau_{i_1\cdots i_n}^{(8)}$ by
\begin{align}
\begin{split}
 &\Co \tau_{i_1\cdots i_n}^{(8)}
 \equiv \frac{1}{(8-n)!}\,\tau_{k_1\cdots k_{8-n}i_1\cdots i_n,\,i_1\cdots i_n}^{(8)}\,
 \rmd x^{k_1}\wedge \cdots \wedge \rmd x^{k_{8-n}}\,,
\\
 &\text{or }\quad \bigl(\Co \tau_{i_1\cdots i_n}^{(8)}\bigr)_{k_1\cdots k_{8-n}}\equiv \tau_{k_1\cdots k_{8-n}i_1\cdots i_n,\,i_1\cdots i_n}^{(8)}\,,
\end{split}
\end{align}
equation \eqref{eq:beta-8} ensures that $\Co \tilde{\beta}^{(8)}_{ij}$ is a (2-tensor-valued) 6-form. 
Further, the relation
\begin{align}
 \tilde{\beta}^{(8)}_{\bfp\bfq} = \frac{1}{2}\,\Co \tilde{\beta}^{(8)}_{\bfp\bfq}\wedge \rmd x^\bfp\wedge \rmd x^\bfq\,,
\label{eq:contracted-beta8}
\end{align}
is shown to be satisfied.

\subsection{$5^2_2$-brane as a source of the $Q$-flux}

By using the fundamental fields of the $\beta$-supergravity, 
the $5^2_2$ background can be written as
\begin{align}
\begin{split}
 \underline{\text{ $5^2_2$ :}}\quad 
 &\rmd \tilde{s}^2 = \rho_2\,\abs{f}^2\,\rmd z\,\rmd \bar{z} + \rmd x^2_{034567} + \rho_2^{-1}\, \rmd x_{89}^2 \,, \quad
 \Exp{2\tilde{\phi}} = \rho_2^{-1} \,, 
\\
 &\tilde{\beta}^{89}= -\rho_1 \,, \quad
 \tilde{\beta}_{89}^{(8)} = \rho_2^{-1}\,\rmd t\wedge \rmd x^3\wedge \cdots \wedge \rmd x^9\,,
\end{split}
\label{eq:522-beta}
\end{align}
which indeed satisfies the equations of motion for the $\beta$-supergravity \cite{Hassler:2013wsa,Andriot:2014uda}. 
This background is quite similar to the NS5-brane background \eqref{eq:NS5}, 
although the $8$-$9$ components of the metric and the dilaton are inverted. 
Since the monodromy around the center simply becomes a gauge transformation ($\tilde{\beta}^{89}\to \tilde{\beta}^{89}-\sigma$), 
called the $\beta$-transformation, 
we can conclude that the $\beta$-supergravity can describe the $5^2_2$ background globally.%%%%%%
\footnote{In fact, the background fields include a cutoff radius $r_\text{c}$ and the geometry gives a good description only for $r\ll r_\text{c}$. 
However, we can smoothly extend the geometry beyond the cutoff radius by introducing additional defect branes, 
which makes the total energy finite \cite{Greene:1989ya,Bergshoeff:2006jj},
and then the cutoff radius $r_\text{c}$ can be interpreted as the distance between the $5^2_2$-brane and a neighboring defect brane \cite{Kikuchi:2012za}.} 
%%%%%%%%%%%%%%%%%%%%%%%%%%%%%%%%%%%%%%%%%%%%%%%%%%%%%%%%%%%%%%%%%%%%%%%%%%%%%%%%%%%%%%%%%%%%%%%%%%%%%%%%%%%%%
On the other hand, the NS5 background, 
obtained from \eqref{eq:522-beta} with the replacements \eqref{eq:exotic-dual}, 
is not single-valued and non-geometric. 
Namely, the $\beta$-supergravity can be considered as a reformulation of the usual supergravity 
which is suitable for a global description of the $5^2_2$ background, instead of the NS5 background. 
Similarly, all effective actions proposed in this paper 
are suitable for a global description of an exotic-brane background, instead of a standard brane background. 

Now, in order to discuss the coupling of the mixed-symmetry tensor 
$\tilde{\beta}^{(8)}_{ij}$ to the $5^2_2$-brane, 
we comment on the relation between the definition of $\tilde{\beta}^{ij}$ or $\tilde{\beta}^{(8)}_{ij}$ 
and that of the corresponding one, $\tilde{B}^{mn}$ or $\tilde{B}_{8}^{mn}$, 
introduced in the study of the effective worldvolume theory of 
the $5^2_2$-brane \cite{Chatzistavrakidis:2013jqa}. 
In \cite{Chatzistavrakidis:2013jqa}, the $2\times 2$-matrix $\tilde{B}^{mn}$ is defined by
\begin{align}
 \tilde{B}^{mn}\equiv \frac{\det B_{mn}}{\det E_{mn}}\,(B^{-1})^{mn} \quad (m,n=8,9)\,,
\end{align}
which is obtained by applying a double $T$-duality $T_{89}$ to $B_{mn}$. 
On the other hand, $\tilde{\beta}^{ij}$, defined in \eqref{eq:beta-fields}, is obtained by applying the $T$-dualities in all spacetime dimensions to $B_{ij}$. 
In order to compare these quantities, 
we assume that the background satisfies $G_{\mu m}=B_{\mu i}=0$ ($\mu,\nu=0,\dotsc, 7$, $m,n=8,9$). 
In this case, we can easily show that the $8$-$9$ components of $\tilde{\beta}^{ij}$ 
coincide with $\tilde{B}^{mn}$\,. 
Moreover, with the same assumptions, the definition of $\tilde{B}_8^{mn}$ 
(see (5.14) of \cite{Chatzistavrakidis:2013jqa}) becomes%%%%%%%%%%%%%%%%%%%%%%%%%%%%%%%%%%%%%%%%%%%%%%
\footnote{Under the assumption $G_{\mu m}=B_{\mu i}=0$, $\tilde{H}^m$ and $\theta_m$ appearing in (5.14) of \cite{Chatzistavrakidis:2013jqa} vanish. 
In addition, $\det(G_{np} + B_{np})$ in the same equation should be corrected as $(\det E_{mn})^2/(\det G_{mn})$\,.}
%%%%%%%%%%%%%%%%%%%%%%%%%%%%%%%%%%%%%%%%%%%%%%%%%%%%%%%%%%%%%%%%%%%%%%%%%%%%%%%%%%%%%%%%%%%%%%%%%%%%%%
\begin{align}
 \rmd \iota_9\iota_8\tilde{B}_{8}^{89} 
 = \Exp{-2\phi} \sqrt{\det G_{mn}}\, \frac{(\det E_{mn})^2}{\det G_{mn}}\, \hat{\star}\,\rmd\tilde{B}^{89} \,, 
\end{align}
where $\hat{\star}$ is the Hodge star operator associated with the metric $G_{\mu\nu}$\,. 
Using that $\tilde{g}_{ij}$ and $\Exp{-2\tilde{\phi}}$ are now given by
\begin{align}
 (\tilde{g}_{ij})=\bpm G_{\mu\nu} & 0 \\ 0 & \frac{\det E_{mn}}{\det G_{mn}}\,G_{mn} \epm \,,\quad 
 \Exp{-2\tilde{\phi}}\sqrt{\det \tilde{g}_{mn}} = \Exp{-2\phi}\sqrt{\det G_{mn}}\,,
\end{align}
and using the isometries in the $89$-directions, we obtain
\begin{align}
 \rmd \tilde{B}_{8}^{89} 
 = \Exp{-2\tilde{\phi}} \tilde{g}_{8k}\, \tilde{g}_{9l}\, \tilde{*}\,\rmd \tilde{B}^{kl} \,, 
\end{align}
which coincides with the definition of $\tilde{\beta}^{(8)}_{89}$ given in \eqref{eq:beta-Bianchi} 
with the identification $\tilde{B}^{89}=\tilde{\beta}^{89}$\,. 
For a general case where $G_{\mu m}$ and $B_{\mu i}$ are no longer assumed to vanish, 
definitions of $\tilde{\beta}^{ij}$ and $\tilde{B}^{mn}$ are different. 
However, this is not a problem since the relation between the field $\beta^{ij}$ 
which magnetically couples to the exotic $5^2_2$-brane, 
and the original background fields $(G_{ij},\,B_{ij})$ depends on a choice of the duality frame. 
Namely, under a duality transformation 
\begin{align}
 \{G_{ij},\,B_{ij},\,\cdots\}
 \to\bigl\{G'_{ij}(G_{ij},\,B_{ij},\,\cdots),\,B'_{ij}(G_{ij},\,B_{ij},\,\cdots),\,\cdots\bigr\}\,,
\end{align}
the relation between $\beta^{ij}$ and the original fields is changed; 
$\beta^{ij}=\tilde{\beta}^{ij}(G_{ij},\,B_{ij})=\widehat{\beta}^{ij}(G'_{ij},\,B'_{ij})$\,. 
In the following, we will basically use the relation 
$\beta^{ij}=\tilde{\beta}^{ij}(G_{ij},\,B_{ij})$ and omit the tilde, 
since with this definition, we can write down a covariant action \eqref{eq:beta-action} 
for $\beta^{ij}$ with all $i$-$j$ components. 

As shown in \cite{Chatzistavrakidis:2013jqa,Kimura:2014upa}, 
the Wess-Zumino term of the $5^2_2(34567,89)$-brane action (smeared in the isometry directions, $x^8$ and $x^9$) can be written as
\begin{align}
\begin{split}
 S_{\WZ}^{5^2_2}
 &= -\mu_{5^2_2}\,n^{89}\,\int_{\cM_6\times T^2_{89}}\Co\beta_{89}^{(8)}\wedge \frac{\rmd x^8\wedge\rmd x^9}{(2\pi R_8)(2\pi R_9)}
 = -\frac{\mu_{5^2_2}\,n^{89}}{(2\pi R_8)(2\pi R_9)}\,\int_{\cM_6\times T^2_{89}} \beta_{89}^{(8)} 
\\
 &= -\mu_{5^2_2}\,\int \beta_{89}^{(8)}\wedge \bdelta^{89}(x-X(\xi)) \quad \bigl(n^{89}\text{: number of the $5^2_2(34567,89)$-branes}\bigr)
\\
 &\Bigl(\bdelta^{\bfp_1\cdots\bfp_n}(x-X(\xi))
  \equiv \frac{n^{\bfp_1\cdots\bfp_n}\,\delta^2(x-X(\xi))}{(2\pi R_{\bfp_1})\cdots(2\pi R_{\bfp_n})}\,\rmd x^1\wedge\rmd x^2\,,\quad n^{\bfp_1\cdots\bfp_n}\in\lZ \Bigr)\,,
\end{split}
\label{eq:SWZ522}
\end{align}
where we used \eqref{eq:contracted-beta8} and $\cM_6$ is the worldvolume of the $5^2_2$-brane, 
and the Ramond-Ramond fields and the worldvolume gauge fields are turned off for simplicity. 
Now, let us consider the dual action which is equivalent to \eqref{eq:beta-action-reduced}, 
and additionally includes the Wess-Zumino term:
\begin{align}
\begin{split}
 S\bigl[\,\tilde{g}_{ij},\,\tilde{\phi},\,\beta^{(8)}_{ij}\,\bigr] 
 &= \frac{1}{2\kappa_{10}^2}\, \int \Bigl[ \Exp{-2\tilde{\phi}}\,\bigl(\tilde{*}\,\tilde{R} + 4\,\rmd\tilde{\phi}\wedge \tilde{*}\,\rmd\tilde{\phi} \bigr)
  -\frac{1}{4}\,\Exp{2\tilde{\phi}}\,\tilde{g}^{ik}\,\tilde{g}^{jl}\,\bQ^{(9)}_{ij}\wedge \tilde{*}\,\bQ^{(9)}_{kl}\Bigr]
\\ 
 &\quad - \mu_{5^2_2}\,\int \frac{1}{2}\,\beta_{\bfp\bfq}^{(8)}\wedge\bdelta^{\bfp\bfq}(x-X(\xi)) \,.
\end{split}
\end{align}
Taking a variation with respect to $\beta_{\bfp\bfq}^{(8)}$, 
we obtain the following equation of motion:
\begin{align}
 \frac{1}{2\kappa_{10}^2}\,\rmd \bQ^{(1)\,\bfp\bfq} 
 = \frac{\mu_{5^2_2}}{(2\pi R_{\bfp})(2\pi R_{\bfq})}\,n^{\bfp\bfq} \,\delta^2(x-X(\xi))\,\rmd x^1\wedge\rmd x^2 \,. 
\label{eq:BI-522}
\end{align}
See \cite{Chatzistavrakidis:2013jqa,Andriot:2014uda} and Appendix D of \cite{Okada:2014wma} 
for the above Bianchi identity for the $Q$-flux in the presence of the $5^2_2$-branes (or the $Q$-branes). 
From \eqref{eq:BI-522}, we conclude that the current for the $5^2_2(n_1\cdots n_5,m_1m_2)$-brane 
(in the absence of the Ramond-Ramond fields) is given by \cite{Okada:2014wma}
\begin{align}
 \tilde{*}\,j_{5^2_2(n_1\cdots n_5,\,m_1m_2)}
 = \frac{(2\pi R_{m_1})(2\pi R_{m_2})}{2\kappa_{10}^2\,\mu_{5^2_2}}\,\rmd \bQ^{(1)\,m_1m_2} \,. 
\end{align}

Having identified the $5^2_2$-brane as a magnetic source of the $Q$-flux, 
it is then natural to investigate an object which electrically couples to the $\beta$-field. 
In the $\beta$-supergravity, such an object will be described by the following (Euclidean) solution:
\begin{align}
\begin{split}
 \rmd \tilde{s}^2 &= \abs{f}^2\,\rmd z\,\rmd \bar{z} + \rmd \tau^2 + \rmd x^2_{34567} + \rho_2\,\rmd x_{89}^2 \,,\quad 
 \Exp{2\tilde{\phi}} = \rho_2 \,, 
\\
 \beta^{(8)}_{89}&= -\rho_1\,\rmd \tau\wedge\rmd x^3\wedge\cdots\wedge\rmd x^9 \,,\quad 
 \beta^{89} = \rho_2^{-1}\,.
\end{split}
\label{eq:beta-2-instanton}
\end{align}
Indeed, this background has a monodromy given by the shift in the mixed-symmetry tensor, 
$\beta^{(8)}_{03\cdots 9,89}\to \beta^{(8)}_{03\cdots 9,89} - \sigma$, 
which gives a non-zero electric charge associated with the $Q$-flux;
\begin{align}
 \int \rmd \beta^{(8)}_{03\cdots 9,89} = -\sigma\,(2\pi R_0)\,(2\pi R_3)\cdots (2\pi R_9) \,.
\end{align}
Since $\beta^{89}$ is a (bi-vector-valued) 0-form, the object which electrically couples to $\beta^{89}$ will be an instanton 
that has two special isometry directions, $x^8$ and $x^9$. 
See section \ref{sec:p=7} for further details about instanton backgrounds. 

Now, let us consider how the $\FF1$ background (or the $1^6_4$ background) can be described in the $\beta$-supergravity. 
If we use the relation \eqref{eq:beta-fields}, 
we will notice that, apparently, we cannot express the background \eqref{eq:F1-BG} (or \eqref{eq:1^6_4-BG}) 
in terms of the fundamental fields of the $\beta$-supergravity 
since the matrix $E_{ij}$ is not invertible. 
However, if we consider a gauge transformation $B_{03}\to B_{03}-a$ ($a$: constant) 
in the defect $\FF1(3)$ background \eqref{eq:F1-BG}, we can calculate $(\tilde{g}_{ij},\,\tilde{\phi},\,\beta^{ij})$\,:
\begin{align}
\begin{split}
 \underline{\text{ $\FF1$ :}}\quad 
 &\rmd\tilde{s}^2= \abs{f}^2\,\rmd z\,\rmd\bar{z} + \Exp{2\tilde{\phi}}\,\bigl(\rmd t^2 - \rmd x_3^2\bigr) + \rmd x_{4\cdots 9}^2 \,, \quad
 \Exp{2\tilde{\phi}}= a^2\,(2\,a^{-1}+ \rho_2)\,,
\\
 &\beta^{(8)}_{03}= - a^2\,\rho_1\,\rmd t\wedge \rmd x^3\wedge\cdots\wedge \rmd x^9\,,\quad 
  \beta^{03} = \frac{a^{-2}}{2\,a^{-1}+ \rho_2} - a^{-1}\,.
\end{split}
\end{align}
If we choose $a=1$ and interchange $t$ with $x^3$, and perform a shift in the cutoff radius $r_\text{c}$ which makes $2+ \rho_2\to \rho_2$, 
the $\FF1(3)$ background becomes
\begin{align}
\begin{split}
 \underline{\text{ $\FF1$ :}}\quad 
 &\rmd\tilde{s}^2= \abs{f}^2\,\rmd z\,\rmd\bar{z} + \Exp{2\tilde{\phi}}\,\bigl(-\rmd t^2 + \rmd x_3^2\bigr) + \rmd x_{4\cdots 9}^2 \,, \quad
 \Exp{2\tilde{\phi}}= \rho_2 \,,
\\
 &\beta^{(8)}_{03}= \rho_1\,\rmd t\wedge \rmd x^3\wedge\cdots\wedge \rmd x^9 \,,\quad 
 \beta^{03} = -\rho_2^{-1} + 1\,,
\end{split}
\end{align}
which is globally well-defined. 
Similarly, the $1^6_4$ background becomes
\begin{align}
\begin{split}
 \underline{\text{ $1^6_4$ :}}\quad 
 &\rmd\tilde{s}^2= \abs{\rho}^2\,\abs{f}^2\,\rmd z\,\rmd\bar{z} + \Exp{2\tilde{\phi}}\,\bigl(-\rmd t^2 + \rmd x_3^2\bigr) + \rmd x_{4\cdots 9}^2 \,, \quad 
 \Exp{2\tilde{\phi}}= \frac{\rho_2}{\abs{\rho}^2} \,, 
\\
 &\beta^{(8)}_{03}= -\frac{\rho_1}{\abs{\rho}^2}\,\rmd t\wedge \rmd x^3\wedge\cdots\wedge \rmd x^9 \,,\quad 
 \beta^{03} = -\frac{\abs{\rho}^2}{\rho_2} + 1\,,
\end{split}
\end{align}
which is not single-valued as is the case with the original background \eqref{eq:1^6_4-BG}. 
A possible way to globally describe the $1^6_4$ background is discussed in section \ref{sec:NS-NS}. 

\section{Ramond-Ramond counterpart of the $\beta$-supergravity}
\label{sec:S-dual-bega}

In the previous section, we reviewed the $\beta$-supergravity 
and explained that it is suitable for describing the non-geometric $5^2_2$ background. 
If we perform an $S$-duality, the $5^2_2$-brane is mapped to the D5$_2$-brane (or the $5^2_3$-brane). 
At the same time, the non-geometric $Q$-flux, sourced by the $5^2_2$-brane, 
will be mapped to another non-geometric flux, called the $P$-flux, 
which is related to the Ramond-Ramond 2-form $C^{(2)}$ instead of the $B$-field. 
In this section, we write down an effective action for the $P$-flux 
utilizing the techniques of the $\beta$-supergravity 
and examine the $\DD5_2$-brane as a magnetic source of the $P$-flux. 

\subsection{The effective action for the $P$-flux}

Let us begin with the type IIB action in the ten-dimensional Einstein frame:
\begin{align}
\begin{split}
 S&= \frac{1}{2\kappa_{10}^2}\,
 \int \Bigl[ *_{\EE} R_{\EE} - \frac{\rmd\tau\wedge *_{\EE}\rmd\overline{\tau}}{2\,(\im\tau)^2}
\\
 &\quad\qquad\qquad - \frac{1}{2\im\tau}\,\bigl(\rmd C^{(2)}-\tau\,\rmd B^{(2)}\bigr)\wedge *_{\EE}\bigl(\rmd C^{(2)}-\overline{\tau}\,\rmd B^{(2)}\bigr)
  -\frac{1}{4}\,F^{(5)}\wedge *_{\EE} F^{(5)}\Bigr] \,,
\end{split}
\label{eq:SIIB}
\end{align}
where the Chern-Simons term is dropped since it is irrelevant for the following discussions, 
and we have defined the axio-dilaton $\tau$ by
\begin{align}
 \tau \equiv C^{(0)} + \ii\Exp{-\phi}\,.
\end{align}
If we redefine the axio-dilaton by
\begin{align}
 \widetilde{\tau} \equiv \widetilde{C}^{(0)} +\ii \Exp{- \widecheck{\phi}} 
 \equiv -1/\tau = \frac{-C^{(0)}+ \ii \Exp{-\phi}}{\abs{\tau}^2} \,,
\end{align}
the action becomes
\begin{align}
\begin{split}
 S &= \frac{1}{2\kappa_{10}^2}\,
 \int \Bigl[ *_{\EE} R_{\EE} - \frac{\rmd \widetilde{\tau}\wedge *_{\EE}\rmd\overline{\widetilde{\tau}}}{2\,(\im\widetilde{\tau})^2}
\\
 &\quad\qquad\qquad - \frac{1}{2\im\widetilde{\tau}}\,\bigl(\rmd B^{(2)}-\widetilde{\tau}\,\rmd C^{(2)}\bigr)
  \wedge *_{\EE}\bigl(\rmd B^{(2)}-\overline{\widetilde{\tau}}\,\rmd C^{(2)}\bigr)
          -\frac{1}{4}\,F^{(5)}\wedge *_{\EE} F^{(5)}\Bigr] \,.
\end{split}
\end{align}
Then, in the ``string frame'' given by $\widecheck{g}_{ij} \equiv \Exp{\widecheck{\phi}/2}\,G^{\EE}_{ij} 
= \Exp{\phi/2}\,\abs{\tau}\,G^{\EE}_{ij}= \abs{\tau}\,G_{ij}$\,, 
the action can be rewritten as
\begin{align}
\begin{split}
 S &= \frac{1}{2\kappa_{10}^2}\,
    \int \Exp{-2\widecheck{\phi}}\,\bigl(\widecheck{*}\,\widecheck{R} + 4\,\rmd\widecheck{\phi}\wedge \widecheck{*}\,\rmd\widecheck{\phi}
     - \frac{1}{2}\,\rmd C^{(2)}\wedge \widecheck{*}\,\rmd C^{(2)}\bigr)
\\
 &\quad -\frac{1}{4\kappa_{10}^2}\,
    \int \bigl[\rmd \widetilde{C}^{(0)}\wedge \widecheck{*}\,\rmd \widetilde{C}^{(0)}
    + \bigl(\rmd B^{(2)}-\widetilde{C}^{(0)}\,\rmd C^{(2)}\bigr)\wedge \widecheck{*}\,\bigl(\rmd B^{(2)}-\widetilde{C}^{(0)}\,\rmd C^{(2)}\bigr)
\\
 &\quad\qquad\qquad\quad + \frac{1}{2}\,F^{(5)}\wedge \widecheck{*}\, F^{(5)} \bigr]\,.
\end{split}
\label{eq:SIIB-NS7}
\end{align}
In a simple case of $B^{(2)}= C^{(0)}= C^{(4)}=0$\,, the above action reduces to
\begin{align}
 S\bigl[\,\widecheck{g}_{ij},\,\widecheck{\phi},\,C^{(2)}\,\bigr] 
 = \frac{1}{2\kappa_{10}^2}\,\int \Exp{-2\widecheck{\phi}}\,\Bigl(\widecheck{*}\,\widecheck{R}
  + 4\,\rmd\widecheck{\phi}\wedge \widecheck{*}\,\rmd\widecheck{\phi} 
  - \frac{1}{2}\,\rmd C^{(2)}\wedge \widecheck{*}\,\rmd C^{(2)}\Bigr) \,,
\end{align}
which has the same structure with the NS-NS action. 
We can thus use the techniques of the $\beta$-supergravity 
to rewrite the action into the following form:
\begin{align}
\begin{split}
 S\bigl[\,g'_{ij},\,\phi',\,\gamma^{ij}\,\bigr]
 &= \frac{1}{2\kappa_{10}^2}\,\int\rmd^{10}x \sqrt{\abs{g'}}\, \Exp{-2\phi'}\, 
 \Bigl(R'+4g^{\prime\, ij}\,\partial_i \phi\,\partial_j \phi + \widecheck{\cR}(g'_{ij},\,\gamma^{ij}) 
\\
 &\qquad\qquad\qquad\qquad\qquad\qquad - \frac{1}{2}\,\abs{S^{ijk}}^2 
                                       + 4\,\abs{\gamma^{ij}\,\partial_j \phi -\cT^i(g'_{ij},\,\gamma^{ij})}^2\Bigr)\,,
\end{split}
\end{align}
where $g'_{ij}$, $\gamma^{ij}$, $\phi$, and $S^{ijk}$ are defined by
\begin{align}
\begin{split}
 g'_{ij}&\equiv F_{ik}\,F_{jl}\,\widecheck{g}^{kl} \,, \quad
 \gamma^{ij} \equiv \bigl(F^{-1}\bigr)^{ik}\,\bigl(F^{-1}\bigr)^{jl}\,C^{(2)}_{kl}\,,\quad 
 \Exp{-2\phi'}\sqrt{\abs{g'}}\equiv \Exp{-2\widecheck{\phi}}\sqrt{\abs{\widecheck{g}}}\,,
\\
 F&\equiv \bigl(F_{ij}\bigr)\equiv \bigl(\widecheck{g}_{ij} - C^{(2)}_{ij}\bigr) 
   = \bigl(\Exp{-\phi}G_{ij} - C^{(2)}_{ij}\bigr)\,,\quad 
 S^{ijk}\equiv 3\,\gamma^{l[i}\,\partial_l\gamma^{jk]} \,,
\end{split}
\end{align}
and $R'$ is the Ricci scalar associated with $g'_{ij}$\,, 
and $\widecheck{\cR}$ and $\cT^i$ are the same as those defined in \eqref{eq:beta-definitions}. 
We then make a further redefinition
\begin{align}
 \tilde{g}_{ij} \equiv \Exp{-\phi'} g'_{ij}\,,\quad 
 \Exp{\tilde{\phi}}\equiv \Exp{-\phi'} \,,
\end{align}
which brings the action into the following form:
\begin{align}
\begin{split}
 &\tilde{S}\bigl[\,\tilde{g}_{ij},\,\tilde{\phi},\,\gamma^{ij}\,\bigr] 
\\
 &= \frac{1}{2\kappa_{10}^2}\int\rmd^{10}x \sqrt{\abs{\tilde{g}}}\,
 \Bigl[ \Exp{-2\tilde{\phi}} \,\bigl(\tilde{R}+4\,\tilde{g}^{ij}\,\partial_i \phi\,\partial_j \phi \bigr)
      -\frac{1}{2}\,\Exp{-6\tilde{\phi}} \,\abs{S^{ijk}}^2 
\\
 &\qquad\qquad\qquad +\Exp{-4\tilde{\phi}}\,\Bigl(\tilde{g}_{ij}\,\widecheck{\cR}^{ij}(\Exp{-\tilde{\phi}}\tilde{g}_{ij},\,\gamma^{ij})
    + 4\,\absb{\gamma^{ij}\,\partial_j \phi -\cT^i(\Exp{-\tilde{\phi}}\tilde{g}_{ij},\,\gamma^{ij})}^2\Bigr)\Bigr] \,.
\end{split}
\label{eq:gamma-action}
\end{align}

To summarize, we performed the redefinition of fields
\begin{align}
\begin{split}
 &\tilde{g}_{ij}= \frac{\abs{\det (\Exp{-\phi}G_{ij})}^{1/2}}{\abs{\det F_{ij}}^{1/2}}\,\Exp{2\phi} F_{ik}\,F_{jl}\,G^{kl} \,, \quad 
 \gamma^{ij} = \bigl(F^{-1}\bigr)^{ik}\,\bigl(F^{-1}\bigr)^{jl}\,C^{(2)}_{kl}\,,
\\
 &\Exp{2\tilde{\phi}} = \frac{\abs{\det (\Exp{-\phi}G_{ij})}}{\abs{\det F_{ij}}}\,\Exp{2\phi}
\,,\quad
 F_{ij} = \Exp{-\phi} G_{ij} - C^{(2)}_{ij} \,,
\end{split}
\label{eq:gamma-field-redefinition}
\end{align}
and the resulting action \eqref{eq:gamma-action} is equal to the type IIB action with $B^{(2)}= C^{(0)}= C^{(4)}=0$, up to total derivative terms. 
As we expect naturally, the above action can also be obtained by making the following replacements (like the $S$-duality) 
in the action of the $\beta$-supergravity:
\begin{align}
\begin{alignedat}{3}
 G_{ij} &\to \Exp{-\phi}\, G_{ij}\,,\quad&
 \phi &\to - \phi\,,\quad& B_{ij}&\to -C^{(2)}_{ij} \,,
\\
 \tilde{g}_{ij} &\to \Exp{-\tilde{\phi}}\, \tilde{g}_{ij}\,, \quad&
 \tilde{\phi}&\to -\tilde{\phi}\,,\quad& \beta^{ij}&\to -\gamma^{ij}\,.
\label{eq:gen-S-dual}
\end{alignedat}
\end{align}

As in the case of the $\beta$-supergravity, 
if we assume that any derivative $\partial_i$ contracted with $\gamma^{ij}$ vanishes, 
the above action reduces to the following simple action:
\begin{align}
\begin{split}
 \tilde{S}\bigl[\,\tilde{g}_{ij},\,\tilde{\phi},\,\gamma^{ij}\,\bigr] 
 = \frac{1}{2\kappa_{10}^2} \int\Bigl[\,\Exp{-2\tilde{\phi}}\,
   \bigl(\tilde{*}\,\tilde{R} + 4\,\rmd\phi\wedge \tilde{*}\,\rmd\phi\bigr)
  -\frac{1}{4} \Exp{-4\tilde{\phi}}\,\tilde{g}_{ik}\,\tilde{g}_{jl}\,
   \bP^{(1)\,ij}\wedge \tilde{*}\bP^{(1)\,kl}\,\Bigr] \,,
\end{split}
\label{eq:gamma-action-reduced}
\end{align}
where we defined the $P$-flux by 
$\bP^{(1)\,ij}\equiv P_k{}^{ij}\,\rmd x^k \equiv \rmd \gamma^{ij}$. 

\subsection{$5^2_3$-brane as a source of the $P$-flux}

The $\DD 5_2(34567,89)$ background written in the new background fields $(\tilde{g}_{ij},\,\tilde{\phi},\,\gamma^{ij})$ is given by
\begin{align}
\begin{split}
 \underline{\text{ $\DD 5_2$ :}}\quad 
 &\rmd \tilde{s}^2 = \rho_2^{1/2}\, \bigl(\rho_2\,\abs{f}^2\,\rmd z\,\rmd \bar{z}+\rmd x^2_{034567}\bigr) + \rho_2^{-1/2}\, \rmd x^2_{89} \,, \quad
 \Exp{2\tilde{\phi}} = \rho_2 \,,
\\
 &\gamma^{89} = \rho_1 \,, \quad 
 \gamma^{(8)}_{89} = -\rho_2^{-1}\,\rmd t \wedge \rmd x^3\wedge \cdots \wedge \rmd x^9 \,.
\end{split}
\label{eq:523-gamma}
\end{align}
We can confirm that this background indeed satisfies the equations of motion 
derived from the action \eqref{eq:gamma-action-reduced} 
(see \eqref{eq:EOM1}--\eqref{eq:EOM3} for the explicit form). 
Note that this is a geometric background since the monodromy is given by a gauge transformation 
that corresponds to the shift in the $\gamma$-field, $\gamma^{89}\to \gamma^{89}+\sigma$, 
to be called the \emph{$\gamma$-transformation}. 

According to \cite{Chatzistavrakidis:2013jqa,Kimura:2014upa}, 
the Wess-Zumino term of the $5^2_3(34567,89)$-brane action 
(smeared in the isometry directions, $x^8$ and $x^9$) 
is written as
\begin{align}
 S^{5^2_3}_{\WZ} = - \mu_{5^2_3} \,\int \gamma_{89}^{(8)}\wedge\bdelta^{89}(x-X(\xi)) \,,
\end{align}
where the $B$-field, the Ramond-Ramond 0- and 4-forms, and the worldvolume gauge fields 
are turned off for simplicity, and $\bdelta^{89}(x-X(\xi))$ is defined in \eqref{eq:SWZ522}. 
As in the case of the $5^2_2$-brane, if we consider the action
\begin{align}
\begin{split}
 S\bigl[\,\tilde{g}_{ij},\,\tilde{\phi},\,\gamma^{(8)}_{ij}\,\bigr] 
 &= \frac{1}{2\kappa_{10}^2}\, \int \Bigl[\Exp{-2\tilde{\phi}}\,\bigl(\tilde{*} \tilde{R} + 4\,\rmd\tilde{\phi}\wedge \tilde{*}\,\rmd\tilde{\phi} \bigr)
  -\frac{1}{4}\,\Exp{4\tilde{\phi}}\,\tilde{g}^{ik}\,\tilde{g}^{jl}\,\bP^{(9)}_{ij}\wedge \tilde{*}\,\bP^{(9)}_{kl} \Bigr]
\\ 
 &\quad - \mu_{5^2_3}\,\int \frac{1}{2}\, \gamma_{\bfp\bfq}^{(8)}\wedge\bdelta^{\bfp\bfq}(x-X(\xi)) \,,
\end{split}
\end{align}
and take a variation with respect to $\gamma_{\bfp\bfq}^{(8)}$, we obtain the Bianchi identity for the $P$-flux with a source term:
\begin{align}
 \frac{1}{2\kappa_{10}^2}\,\rmd \bP^{(1)\,\bfp\bfq} 
 = \frac{\mu_{5^2_3}}{(2\pi R_\bfp)(2\pi R_\bfq)}\,n^{\bfp\bfq}\, \delta^2(x-X(\xi))\,\rmd x^1\wedge\rmd x^2 \,. 
\end{align}

As in the case of the $\beta$-supergravity, 
we can further find a solution corresponding to the (Euclidean) background of an instanton that couples to $\gamma^{ij}$ electrically. 
The explicit form of the background fields is given later in the next section (see \eqref{eq:gamma-instantons}). 

\section{Effective actions for non-geometric fluxes}
\label{sec:exotic-p-branes}

In this section, we derive the effective actions for various polyvectors $\gamma^{i_1\cdots i_{7-p}}$, 
which are generalizations of the (simple) action \eqref{eq:gamma-action-reduced} 
for $\gamma^{ij}$ to arbitrary $(7-p)$-vectors. 
From these actions, we can also obtain the effective actions for 
the non-geometric $Q$-fluxes; $Q_k{}^{ij}$ and $Q_k{}^{i_1\cdots i_6}$. 
Since the derivation presented here does not rely on the results of the $\beta$-supergravity, 
as for the action for the $Q$-flux, 
the derivation can be regarded as another derivation of the (simplified) $\beta$-supergravity action. 
Further, we find two kinds of solutions which correspond to 
the exotic-brane backgrounds and their electric duals. 

\subsection{Effective actions for the $P$-fluxes}
\label{sec:action-P-flux}

In this subsection, we consider the following ansatz, 
which is crucial in deriving the effective actions for the non-geometric $P$-fluxes:
\begin{align}
\begin{split}
 &\rmd s^2=G_{\bfa\bfb}\,\rmd x^\bfa\,\rmd x^\bfb+G_{\bfp\bfq}\,\rmd x^\bfp\,\rmd x^\bfq \qquad (\,\text{i.e.~}G_{\bfa\bfp}=0 \,)\,, 
\\
 &B^{(2)}=0\,,\quad 
 C^{(p+1)}_{\bfa i_1\cdots i_p}=0\,,\quad \text{isometries in the $x^\bfp$-directions}\,.
\label{eq:assumptions}
\end{split}
\end{align}
It is important to note that, throughout this paper, 
the bold indices $\bfa,\bfb,\bfc$ run only over $1,2$, 
while $\bfp,\dotsc,\bft$ run over $0,3,\dotsc,9$. 
Now, due to a technical reason explained below, 
we start from the type II$^\star$ theory \cite{Hull:1998vg}. 
Since the $B$-field is assumed to vanish, the action is given by
\begin{align}
 S = \frac{1}{2\kappa_{10}^2}\,
    \int \Bigl[\Exp{-2\phi}\,\bigl(*\,R + 4\,\rmd \phi\wedge *\,\rmd \phi\bigr)
  + \sum_p \frac{a_p}{2}\, \rmd C^{(p+1)}\wedge *\,\rmd C^{(p+1)}\Bigr] \,,
\label{eq:typeII*}
\end{align}
where $p$ is summed over $0,2$ for the type IIA$^\star$ theory 
while $-1,1,3$ for the type IIB$^\star$ theory, 
and $a_p$ are constants given by $a_p=1$ ($p=-1,0,1,2$) and $a_3=1/2$. 

Let us introduce $(7-p)$-vector fields $\gamma^{i_1\cdots i_{7-p}}$ by
\begin{align}
\begin{split}
 &\gamma^{\bfs_1\cdots \bfs_{7-p}}
 \equiv \frac{1}{(p+1)!}\,\epsilon^{\bfq_{p+1}\cdots \bfq_1\bfs_1\cdots \bfs_{7-p}}\,C^{(p+1)}_{\bfq_1\cdots \bfq_{p+1}} \,, 
\\
 &\gamma^{\bfa i_1\cdots i_{6-p}}\equiv 0 \qquad 
 \bigl(\epsilon^{\bfp_1\cdots \bfp_8}=\epsilon^{[\bfp_1\cdots \bfp_8]}\,,\ \epsilon^{03\cdots 9}=1\bigr) \,. 
\end{split}
\label{eq:gamma-def}
\end{align}
Note that, if the metric is diagonal, 
the map $C^{(p+1)}_{\bfq_1\cdots \bfq_{p+1}}\to \gamma^{\bfs_1\cdots \bfs_{7-p}}$ 
corresponds to performing eight $T$-dualities, $T_{03\cdots 9}$, 
which includes a timelike $T$-duality.%%
\footnote{%%%%%%%%%%%%%%%%%%%%%%%%%%%%%%%%%%%%%%%%%%%%%%%%%%%%%%%%%%%%%%%%%%%%%%%%%%%%%%%%%
If we instead define $\gamma$ as $C^{(p+1)}_{j_1\cdots j_{p+1}}= \epsilon_{j_{p+1}\cdots j_1i_1\cdots i_{9-p}}\,\gamma^{i_1\cdots i_{9-p}}/(9-p)!$\,, 
which corresponds to taking $T$-dualities in all spacetime directions, $\gamma^{i_1\cdots i_{9-p}}$ coincides with $\tilde{C}^{i_1\cdots i_{9-p}}$ 
defined in (6.15) of \cite{Hohm:2011dv}.} 
%%%%%%%%%%%%%%%%%%%%%%%%%%%%%%%%%%%%%%%%%%%%%%%%%%%%%%%%%%%%%%%%%%%%%%%%%%%%%%%%%%%%%%%%%%
Thus, in order to make $\gamma^{i_1\cdots i_{7-p}}$ into a background field in the type II theory, 
$C^{(p+1)}$ should be a field in the type II$^\star$ theory, 
and this is the reason why we start from the type II$^\star$ theory. 
In the following, we rewrite the action \eqref{eq:typeII*} regarding $\gamma^{i_1\cdots i_{7-p}}$ 
as a fundamental variable, instead of the form field $C^{(p+1)}$\,. 
Under the assumptions \eqref{eq:assumptions}, 
the Ramond-Ramond part of the action \eqref{eq:typeII*} becomes
\begin{align}
 \frac{1}{2\kappa_{10}^2}\int\rmd^{10}x \sqrt{\abs{G}}\,\sum_p\frac{a_p}{2\,(p+1)!}\,G^{\bfa\bfb}\,G^{\bfp_1\bfq_1}\cdots G^{\bfp_{p+1}\bfq_{p+1}}\,
 \partial_\bfa C^{(p+1)}_{\bfp_1\cdots \bfp_{p+1}}\,\partial_\bfb C^{(p+1)}_{\bfq_1\cdots \bfq_{p+1}} \,.
\end{align}
By substituting the relation
\begin{align}
 C^{(p+1)}_{\bfq_1\cdots \bfq_{p+1}}
 = \frac{1}{(7-p)!}\,\epsilon_{\bfq_{p+1}\cdots \bfq_1\bfs_1\cdots \bfs_{7-p}}\,\gamma^{\bfs_1\cdots \bfs_{7-p}} \quad
 \bigl(\epsilon_{\bfp_1\cdots \bfp_8}=\epsilon_{[\bfp_1\cdots \bfp_8]}\,,\ \epsilon_{03\cdots 9}=1\bigr)\,,
\end{align}
the action can be written as
\begin{align}
 -\frac{1}{2\kappa_{10}^2}\,\int\rmd^{10}x\sqrt{\abs{G}}\,\sum_p\frac{a_p\,\Delta^{-1}}{2\,(7-p)!}\,G^{\bfa\bfb}\,G_{\bfr_1\bfs_1}\cdots G_{\bfr_{7-p}\bfs_{7-p}}\,\partial_\bfa \gamma^{\bfr_1\cdots \bfr_{7-p}}\,\partial_\bfb \gamma^{\bfs_1\cdots \bfs_{7-p}} \,,
\end{align}
where we defined $\Delta\equiv \abs{\det G_{\bfp\bfq}}$ and used the identity
\begin{align}
\begin{split}
 &\frac{1}{(p+1)!\,[\,(7-p)!\,]^2}\,G^{\bfp_1\bfq_1}\cdots G^{\bfp_{p+1}\bfq_{p+1}}\,\epsilon_{\bfp_1\cdots \bfp_{p+1}\bfr_1\cdots \bfr_{7-p}}\,\epsilon_{\bfq_1\cdots \bfq_{p+1}\bfs_1\cdots \bfs_{7-p}}
\\
 &= -\frac{\Delta^{-1}}{(7-p)!}\, G_{\bft_1\bfs_1}\cdots G_{\bft_{7-p}\bfs_{7-p}}\,
    \delta^{[\bft_1}_{[\bfr_1}\cdots \delta^{\bft_{7-p}]}_{\bfr_{7-p}]}\,.
\end{split}
\end{align}
Thus, the action becomes
\begin{align}
 S = \frac{1}{2\kappa_{10}^2}\,
    \int \biggl[\Exp{-2\phi}\,\bigl(*\,R + 4\,\rmd \phi\wedge *\,\rmd \phi\bigr)
 -\sum_p \frac{a_p\,\Delta^{-1}}{2\,(7-p)!}\,\bP^{(1)\,i_1\cdots i_{7-p}}\wedge *\,\bP^{(1)}_{i_1\cdots i_{7-p}}\biggr] \,,
\label{eq:RR-action}
\end{align}
where we defined the $P$-flux by
\begin{align}
 \bP^{(1)\, i_1\cdots i_{7-p}} \equiv P_j{}^{i_1\cdots i_{7-p}}\,\rmd x^j \equiv \rmd \gamma^{i_1\cdots i_{7-p}}\,.
\end{align}
Due to the presence of $\Delta^{-1}$, 
if we treat $\bP^{(1)\,i_1\cdots i_{7-p}}$ as tensors (see footnote \ref{foot:tensor})
this action is not invariant under diffeomorphisms in the eight-dimensional spacetime, spanned by $x^\bfp$. 
This issue is resolved by making a redefinition of the metric and the dilaton. 

Since the redefinition depends on the degree $p$, 
in the following, we consider the case where only a Ramond-Ramond $p$-form (with $p\neq 3$) is non-vanishing:
\begin{align}
 S = \frac{1}{2\kappa_{10}^2}\,
    \int \biggl[\Exp{-2\phi}\,\bigl(*\,R + 4\,\rmd \phi\wedge *\,\rmd \phi\bigr)
 - \frac{\Delta^{-1}}{2\,(7-p)!}\,\bP^{(1)\,i_1\cdots i_{7-p}}\wedge *\,\bP^{(1)}_{i_1\cdots i_{7-p}}\biggr] \,.
\label{eq:p-vector-action}
\end{align}
In this case, with the redefinition of the metric and the dilaton
\begin{align}
\begin{split}
 \rmd\tilde{s}^2 
 &\equiv \Exp{\frac{4}{p-3}\,\phi} \Exp{\frac{p+1}{p-3}\,\eta} G_{\bfa\bfb}\,\rmd x^\bfa\,\rmd x^\bfb + \Exp{\eta} G_{\bfp\bfq}\,\rmd x^\bfp\,\rmd x^\bfq\,,
\\
 \Exp{2\tilde{\phi}}&\equiv \Exp{2\phi+4\eta}\,,\quad 
 \Exp{\eta}\equiv \bigl(\Exp{-2\phi}\Delta^{1/2}\bigr)^{\frac{2}{p-3}} \,,
\label{eq:gamma-p-redefinitions}
\end{split}
\end{align}
we can show that the action \eqref{eq:p-vector-action} 
is equal to the following action up to total derivative terms (see appendix \ref{app:derivation} for the detail):
\begin{align}
 S = \frac{1}{2\kappa_{10}^2}\,
    \int \biggl[\Exp{-2\tilde{\phi}}\, \bigl(\tilde{*}\,\tilde{R} + 4\,\rmd \tilde{\phi}\wedge \tilde{*}\,\rmd \tilde{\phi}\bigr)
 - \frac{1}{2\,(7-p)!}\,\Exp{-4\tilde{\phi}}\bP^{(1)\,i_1\cdots i_{7-p}}\wedge \tilde{*}\,\bP^{(1)}_{i_1\cdots i_{7-p}}\biggr] \,,
\label{eq:gamma-action-p}
\end{align}
which reduces to the action \eqref{eq:gamma-action-reduced} in a special case of $p=5$\,. 

The equations of motion are obtained as follows:
\begin{align}
 &\tilde{R}+4\bigl(\tilde{\nabla}^i\partial_i\tilde{\phi}-\tilde{g}^{ij}\,\partial_i\tilde{\phi}\,\partial_j\tilde{\phi}\bigr)
 -\frac{\Exp{-2\tilde{\phi}}}{(7-p)!}\,P_i{}^{j_1\cdots j_{7-p}}\,P^i{}_{j_1\cdots j_{7-p}} =0 \,,
\label{eq:EOM1}
\\
 &\tilde{R}_{ij} +2\tilde{\nabla}_i\partial_j\tilde{\phi}-\frac{\Exp{-2\tilde{\phi}}}{2\,(7-p)!}\,
 \Bigl(P_i{}^{k_1\cdots k_{7-p}}\,P_{jk_1\cdots k_{7-p}} -(7-p)\,P_{k_1i}{}^{k_2\cdots k_{7-p}}\,P^{k_1}{}_{jk_2\cdots k_{7-p}}
\nn\\
 &\qquad\qquad\qquad\qquad\qquad\qquad\qquad +\frac{1}{2}\,P_k{}^{l_1\cdots l_{7-p}}\,P^k{}_{l_1\cdots l_{7-p}} \,\tilde{g}_{ij} \Bigr) =0 \,,
\label{eq:EOM2}
\\
 &\rmd \bP^{(9)}_{i_1\cdots i_{7-p}}=0\,,\quad 
 \bP^{(9)}_{i_1\cdots i_{7-p}} \equiv 
 \Exp{-4\tilde{\phi}}\, \tilde{g}_{i_1j_1}\cdots\tilde{g}_{i_{7-p}j_{7-p}}\, \tilde{*}\, \bP^{(1)\,j_1\cdots j_{7-p}}\,.
\label{eq:EOM3}
\end{align}

In this theory, we can find the following two solutions. 
The first one
\begin{align}
\begin{split}
 \underline{\text{ $\DD p_{7-p}$ :}}\quad 
 &\rmd \tilde{s}^2 = \rho_2^{1/2}\,\bigl(\rho_2\,\abs{f}^2\,\rmd z\,\rmd \bar{z} +\rmd x^2_{0n_1\cdots n_p}\bigr) + \rho_2^{-1/2}\, \rmd x^2_{m_1\cdots m_{7-p}} \,, 
\quad
 \Exp{2\tilde{\phi}} = \rho_2^{\frac{p-3}{2}} \,, 
\\
 &\gamma^{m_1\cdots m_{7-p}} = \rho_1 \,, 
\quad
 \gamma^{(8)}_{m_1\cdots m_{7-p}} = -\rho_2^{-1}\,\rmd t \wedge \rmd x^3\wedge \cdots \wedge \rmd x^9 \,,
\end{split}
\label{eq:gamma-Dp_7-p}
\end{align}
is a generalization of \eqref{eq:523-gamma} and has a monodromy given by 
a $\gamma$-transformation, $\gamma^{m_1\cdots m_{7-p}}\to \gamma^{m_1\cdots m_{7-p}} +\sigma$. 
That is, it has a magnetic charge associated with the $P$-flux and should correspond to the background of the exotic $\DD p_{7-p}$-brane. 

On the other hand, the second (Euclidean) solution
\begin{align}
\begin{split}
 &\rmd \tilde{s}^2 = \rho_2^{-1/2}\, \bigl(\abs{f}^2\,\rmd z\,\rmd \bar{z}+\rmd\tau^2+\rmd x^2_{n_1\cdots n_p}\bigr)+ \rho_2^{1/2}\, \rmd x^2_{m_1\cdots m_{7-p}} \,, 
\quad
 \Exp{2\tilde{\phi}} = \rho_2^{\frac{3-p}{2}} \,, 
\\
 &\gamma^{(8)}_{m_1\cdots m_{7-p}} = -\rho_1\,\rmd \tau \wedge \rmd x^3\wedge \cdots \wedge \rmd x^9 \,, \quad 
 \gamma^{m_1\cdots m_{7-p}} = \rho_2^{-1} \,,
\end{split}
\label{eq:gamma-instantons}
\end{align}
has an electric charge associated with the $P$-flux 
and will correspond to an instanton with special isometry directions, $x^{m_1},\dotsc,x^{m_{7-p}}$, 
which is similar to the solution obtained in \eqref{eq:beta-2-instanton}. 

In the case of $p=3$, we cannot perform the redefinition \eqref{eq:gamma-p-redefinitions}, 
and do not know how to derive the action \eqref{eq:gamma-action-p}. 
However, if the action is derived from a suitable field redefinition, 
we can confirm that the above backgrounds with $p=3$ indeed satisfy the equations of motion. 

\subsection{Relating the new and the original background fields}

In this subsection, we discuss the exotic duality further 
and investigate a relation between the fundamental fields of the theory \eqref{eq:gamma-action-p} 
and the standard background fields. 

\paragraph*{\underline{Exotic duality}\\}

In order to examine the exotic duality, let us consider a simple configuration
\begin{align}
\begin{split}
 &\rmd s^2 = G_{\mu\nu}\,\rmd x^\mu \rmd x^\nu+G_{\sfm\sfn}\,\rmd x^\sfm \rmd x^\sfn\quad (G_{\sfm\sfn}:\text{ diagonal}) \,,
\\
 &C^{(7-p)} = C^{(7-p)}_{(p+3)\cdots 9}\,\rmd x^{p+3}\wedge\cdots\wedge\rmd x^9
\\
 &(\mu,\nu=0,\dotsc,p+2\,,\quad \sfm,\sfn=p+3,\dotsc,9)\,,
\end{split}
\label{eq:simple-config}
\end{align}
which includes the $\DD p(3\cdots (p+2))$ and the $\DD p_{7-p}(3\cdots (p+2),(p+3)\cdots 9)$ background in the type II theory. 
For this kind of simple background, it is convenient to take a $T_{9\cdots (p+3)}$-duality and describe it in the type IIB theory. 
The resulting type IIB background (in the Einstein frame) after performing the $T_{9\cdots (p+3)}$-duality is given by
\begin{align}
\begin{split}
 \rmd \tilde{s}_{\EE}^2 
 &= \Exp{-\tilde{\phi}/2}\,\bigl(\tilde{G}_{\mu\nu}\,\rmd x^\mu \rmd x^\nu + \tilde{G}_{\sfm\sfn}\,\rmd x^\sfm \rmd x^\sfn\bigr) \quad 
 \bigl(\tilde{G}_{\mu\nu}\equiv G_{\mu\nu}\,,\ \tilde{G}_{\sfm\sfn}\equiv G^{\sfm\sfn}\bigr)\,,
\\
 \tilde{\tau} &\equiv \tilde{C}^{(0)} +\ii\Exp{-\tilde{\phi}}
 \equiv C^{(7-p)}_{(p+3)\cdots 9} + \ii\sqrt{\det G_{\sfm\sfn}}\, \Exp{-\phi} \equiv \tau_{[(p+3)\cdots 9]}\,.
\end{split}
\end{align}
Since the type IIB theory has the well-known $\SL(2,\lZ)$-duality symmetry, 
\begin{align}
 G^{\EE}_{ij}\to G^{\EE}_{ij}\,,\quad 
 \tau_{[(p+3)\cdots 9]} \to \tau'_{[(p+3)\cdots 9]}
 =\frac{a\,\tau_{[(p+3)\cdots 9]}+b}{c\,\tau_{[(p+3)\cdots 9]}+d}\,,\quad 
  \bpm a & b \\ c & d \epm \in \SL(2,\lZ) \,,
\end{align}
performing the $\SL(2,\lZ)$ transformation followed by a $T_{(p+3)\cdots 9}$-duality transformation, 
we obtain a new background in the original type II theory (see (2.3) of \cite{LozanoTellechea:2000mc} for a similar argument). 
In particular, performing $T_{9\cdots (p+3)}$-$S$-$T_{(p+3)\cdots 9}$-duality transformations, 
the background \eqref{eq:simple-config} is mapped to the following background (in the string frame):
\begin{align}
\begin{split}
 &\rmd s'^2= \abs{\tau_{[(p+3)\cdots 9]}}\,G_{\mu\nu}\,\rmd x^\mu\,\rmd x^\nu
            +\abs{\tau_{[(p+3)\cdots 9]}}^{-1}\,G_{\sfm\sfn}\,\rmd x^\sfm\,\rmd x^\sfn \,,
\\
 &C'^{(7-p)} = -\frac{C^{(7-p)}_{(p+3)\cdots 9}}{\abs{\tau_{[(p+3)\cdots 9]}}^2}\,
               \rmd x^{p+3}\wedge\cdots\wedge\rmd x^9\,,\quad 
  \Exp{2\phi'}=\abs{\tau_{[(p+3)\cdots 9]}}^{p-3}\,\Exp{2\phi} \,.
\end{split}
\label{eq:exotic-special}
\end{align}
For the D$p(3\cdots (p+2))$-brane background, we have $\tau_{[(p+3)\cdots 9]}=\rho_1+\ii\rho_2=\rho(z)$, 
and we can confirm that the above transformation rule coincides with that of the exotic duality transformation 
given in \eqref{eq:exotic-dual}. 

The transformation rule \eqref{eq:exotic-special} under the exotic duality 
can be applied only for $p$-branes extending in the $3\cdots (p+2)$-directions. 
However, for a special case of $p=5$, 
without assuming $C^{(2)} = C^{(2)}_{89}\,\rmd x^{8}\wedge\rmd x^9$, 
we can write down the following transformation rule 
which exchanges a $\DD 5$-brane and an exotic $\DD 5_2$-brane extending in arbitrary $x^\bfp$-directions:%%
\footnote{Note that the transformation rule is deformed in the presence of other background fields, such as the $B$-field.}%%%%%%%%%%%%%%%%%%%%%%%%%%%%%%%%
\begin{align}
\begin{split}
 \widetilde{G}_{\bfa\bfb}&= \abs{\det F_{\bfp\bfq}}^{1/2}\,G_{\bfa\bfb}\,,\quad 
 \widetilde{G}_{\bfp\bfq}= \abs{\det F_{\bfp\bfq}}^{1/2}\,(F^{-1})^{\bfp\bfr}\,(F^{-1})^{\bfq\bfs}\,G_{\bfr\bfs}\,,
\\
 \Exp{2\widetilde{\bphi}}&= \abs{\det F_{\bfp\bfq}}\,\Exp{2\phi}\,, \quad
 \widetilde{C}^{(2)}_{\bfp\bfq} = (F^{-1})^{\bfp\bfr}\,(F^{-1})^{\bfq\bfs}\,C^{(2)}_{\bfr\bfs}\quad 
 \bigl(F_{\bfp\bfq}\equiv \Exp{-\phi}G_{\bfp\bfq}-C^{(2)}_{\bfp\bfq}\bigr)\,. 
\end{split}
\label{eq:exotic-p=5}
\end{align}
This corresponds to an $S$-duality transformation followed by $T_{03\cdots 9}$-dualities and an $S$-duality transformations. 

For general $p$, we do not know a covariant expression like \eqref{eq:exotic-p=5}, 
but we here assume the existence of a transformation rule which interchanges $\DD p$-branes with $\DD p_{7-p}$-branes:
\begin{align}
 \bigl(G_{ij},\,\phi,\,C^{(7-p)}\bigr)
 \overset{\text{exotic duality}}\longrightarrow 
 \bigl(\widetilde{G}_{ij},\,\widetilde{\bphi},\,\widetilde{C}^{(7-p)}\bigr) \,.
\end{align}

Performing an $S$-duality, 
we can also obtain the transformation rule for the fields in the NS-NS sector. 
That is, if we consider a configuration
\begin{align}
\begin{split}
 &\rmd s^2 = G_{\mu\nu}\,\rmd x^\mu \rmd x^\nu+G_{\sfm\sfn}\,\rmd x^\sfm \rmd x^\sfn\quad (G_{\sfm\sfn}:\text{ diagonal}) \,,
\\
 &B^{(7-p)} = B^{(7-p)}_{(p+3)\cdots 9}\,\rmd x^{p+3}\wedge\cdots\wedge\rmd x^9 
\end{split}
\end{align}
with $p=1$ or $p=5$, and define a complex field
\begin{align}
 \tau^{\NS}_{[(p+3)\cdots 9]} \equiv B^{(7-p)}_{(p+3)\cdots 9} + \ii\sqrt{\det G_{\sfm\sfn}}\, \Exp{-\frac{5-p}{2}\,\phi} \,,
\end{align}
the configuration after the action of the exotic duality (i.e.~the $S$-dual of \eqref{eq:exotic-special}) is given by
\begin{align}
\begin{split}
 &\rmd s'^2= \abs{\tau^{\NS}_{[(p+3)\cdots 9]}}^{\frac{p-1}{2}}\,G_{\mu\nu}\,\rmd x^\mu\,\rmd x^\nu 
            +\abs{\tau^{\NS}_{[(p+3)\cdots 9]}}^{\frac{p-5}{2}}\,G_{\sfm\sfn}\,\rmd x^\sfm\,\rmd x^\sfn \,,
\\
 &B'^{(7-p)} = -\frac{B^{(7-p)}_{(p+3)\cdots 9}}{\abs{\tau^{\NS}_{[(p+3)\cdots 9]}}^2}\,
               \rmd x^{p+3}\wedge\cdots\wedge\rmd x^9\,,\quad 
  \Exp{2\phi'}=\abs{\tau^{\NS}_{[(p+3)\cdots 9]}}^{-(p-3)}\,\Exp{2\phi} \,.
\end{split}
\end{align}
We can easily verify that this transformation rule with $p=5$ exchanges the $\NS5(34567)$ background \eqref{eq:NS5} 
and the $5^2_2(34567,89)$ background \eqref{eq:522}, 
and that with $p=1$ exchanges the $\FF1(3)$ background \eqref{eq:F1-BG} 
and the $1^6_4(3,456789)$ background \eqref{eq:1^6_4-BG} with each other. 
Note that the combination $\tau^\NS_{[89]}=B_{89}+\ii\sqrt{\det G_{\sfm\sfn}}$ frequently appears in the discussion of the monodromy 
of NS5(34567)- and $5^2_2(34567,89)$-branes (see e.g., \cite{deBoer:2012ma,Kimura:2014wga}). 

\paragraph*{\underline{On the relation between $\bigl(\tilde{g}_{ij},\,\tilde{\phi},\,\gamma^{i_1\cdots i_{7-p}}\bigr)$ and $\bigl(G_{ij},\,\phi,\,C^{(7-p)}\bigr)$}\\}

Now, we discuss a possible relation between the $\gamma$-field, $\gamma^{i_1\cdots i_{7-p}}$, 
and the standard background fields; $\bigl(G_{ij},\,\phi,\,C^{(7-p)}\bigr)$\,.
We here assume the ansatz \eqref{eq:assumptions}, and further, $G_{pq}$ is diagonal. 
Then, we consider the following sequence of dualities:
\begin{align}
 \bigl(G_{ij},\,\phi,\,C^{(7-p)}\bigr)
 \overset{\text{exotic duality}}\longrightarrow 
 \bigl(\widetilde{G}_{ij},\,\widetilde{\bphi},\,\widetilde{C}^{(7-p)}\bigr) 
 \overset{T_{03456789}}\longrightarrow 
 \bigl(G^\star_{ij},\,\phi^\star,\,C^{\star(p+1)}\bigr)_{\text{II$^\star$}} \,.
\label{eq:dual-fields}
\end{align}
If we consider a non-geometric $\DD p_{7-p}$-brane background on the leftmost side, 
the background in the middle has the same form with the geometric $\DD p$-brane background, 
and the background on the rightmost side has the same form with the E$(7-p)$-brane background \cite{Hull:1998vg} in the type II$^\star$ theory. 
Due to the assumption that $G_{\bfp\bfq}$ is diagonal, the relation between 
$\bigl(\widetilde{G}_{ij},\,\widetilde{\bphi},\,\widetilde{C}^{(7-p)}\bigr)$ 
and 
$\bigl(G^\star_{ij},\,\phi^\star,\,C^{\star(p+1)}\bigr)$ 
is given by
\begin{align}
\begin{split}
 &\widetilde{G}_{\bfa\bfb}=G^\star_{\bfa\bfb}\,,\quad 
  \widetilde{G}^{\bfp\bfq}=G^\star_{\bfp\bfq}\,,\quad 
  \Exp{-2\widetilde{\bphi}}\sqrt{\abs{\det \widetilde{G}_{\bfp\bfq}}}=\Exp{-2\phi^\star}\sqrt{\abs{\det G^\star_{\bfp\bfq}}} \,,
\\
 &\widetilde{C}^{(7-p)}_{\bfs_1\cdots \bfs_{7-p}}
 = \frac{1}{(p+1)!}\,\epsilon^{\bfq_{p+1}\cdots \bfq_1\bfs_1\cdots \bfs_{7-p}}\,C^{\star (p+1)}_{\bfq_1\cdots \bfq_{p+1}} \,.
\end{split}
\end{align}
Then, with the identification of $\bigl(G^\star_{ij},\,\phi^\star,\,C^{\star(p+1)}\bigr)_{\text{II$^\star$}}$ 
and the fields $\bigl(G_{ij},\,\phi,\,C^{(7-p)}\bigr)$ appearing in \eqref{eq:gamma-p-redefinitions}, 
we obtain the following expression for $\bigl(\tilde{g}_{ij},\,\tilde{\phi},\,\gamma^{i_1\cdots i_{7-p}}\bigr)$:
\begin{align}
\begin{split}
 &(\tilde{g}_{ij}) 
 = \begin{pmatrix} \Exp{\frac{4}{p-3}\,\phi^\star} \Exp{\frac{p+1}{p-3}\,\eta} G^\star_{\bfa\bfb}&0\\
 0&\Exp{\eta} G^\star_{\bfp\bfq}
\end{pmatrix} 
 = \begin{pmatrix} \Exp{-\frac{4}{p-3}\,\widetilde{\bphi}}\, \Exp{\frac{7-p}{p-3}\,\eta} \, \widetilde{G}_{\bfa\bfb}&0\\
 0&\Exp{\eta} \widetilde{G}^{\bfp\bfq}
\end{pmatrix} \,,
\\
 &\Exp{2\tilde{\phi}} = \Exp{2\phi^\star+4\eta}=\Exp{-2\widetilde{\bphi}+(7-p)\,\eta} \,,\quad 
 \gamma^{\bfs_1\cdots \bfs_{7-p}}=\widetilde{C}^{(7-p)}_{\bfs_1\cdots \bfs_{7-p}}\,,
\\
 &\Exp{\eta} = \Bigl(\Exp{-2\phi^\star}\sqrt{\abs{\det G^\star_{\bfp\bfq}}}\Bigr)^{\frac{2}{p-3}}
 = \Bigl(\Exp{-2\widetilde{\bphi}}\sqrt{\abs{\det \widetilde{G}_{\bfp\bfq}}}\Bigr)^{\frac{2}{p-3}} \,.
\end{split}
\label{eq:relations}
\end{align}
The inverse relation is given by
\begin{align}
\begin{split}
 &(\widetilde{G}_{ij}) 
  = \begin{pmatrix} \Exp{-\frac{4}{p-3}\,\tilde{\phi}}\, \Exp{\frac{7-p}{p-3}\,\eta}\, \tilde{g}_{\bfa\bfb}&0\\
 0&\Exp{\eta} \tilde{g}^{\bfp\bfq}
\end{pmatrix} \,,\quad
 \Exp{2\widetilde{\bphi}} = \Exp{-2\tilde{\phi}+(7-p)\,\eta} \,,\quad 
\\
 &\widetilde{C}^{(7-p)}_{\bfs_1\cdots \bfs_{7-p}} =\gamma^{\bfs_1\cdots \bfs_{7-p}}\,,
\quad
 \Exp{\eta} = \Bigl(\Exp{-2\tilde{\phi}}\sqrt{\abs{\det \widetilde{g}_{\bfp\bfq}}}\Bigr)^{\frac{2}{p-3}} \,.
\end{split}
\label{eq:inv-relations}
\end{align}

In a case where $\bigl(G_{ij},\,\phi,\,C^{(7-p)}\bigr)$ in \eqref{eq:dual-fields} 
is given by the $\DD p_{7-p}$ background, 
$\bigl(\widetilde{G}_{ij},\,\widetilde{\bphi},\,\widetilde{C}^{(7-p)}\bigr)$ 
in \eqref{eq:dual-fields} 
has the same form with the $\DD p$ background. 
Since the $\DD p$ background \eqref{eq:Dp-brane} satisfies 
$\eta=0$ and $\Exp{-\frac{4}{p-3}\,\widetilde{\bphi}}=\rho_2$,%%%%%%%%%%%%%%%%%%%%%
\footnote{Since $\eta$ is invariant under an $S$-duality or $T$-dualities in the $x^\bfp$-directions, all defect backgrounds 
considered in section \ref{sec:supergravity-description} satisfy $\eta=0$.} 
%%%%%%%%%%%%%%%%%%%%%%%%%%%%%%%%%%%%%%%%%%%%%%%%%%%%%%%%%%%%%%%%%%%%
we obtain the following expression for $\bigl(\tilde{g}_{ij},\,\tilde{\phi},\,\gamma^{i_1\cdots i_{7-p}}\bigr)$:
\begin{align}
\begin{split}
 \underline{\text{ $\DD p_{7-p}$ :}}\quad 
 &\rmd\tilde{s}^2 = \rho_2^{1/2}\,\bigl(\rho_2\,\abs{f}^2\,\rmd z\,\rmd \bar{z} +\rmd x^2_{0n_1\cdots n_p}\bigr) + \rho_2^{-1/2}\, \rmd x^2_{m_1\cdots m_{7-p}}\,,
\\
 &\Exp{2\tilde{\phi}} = \rho_2^{\frac{p-3}{2}} \,, \quad \gamma^{m_1\cdots m_{7-p}} = \rho_1 \,,
\end{split}
\end{align}
which exactly reproduces the solution found in \eqref{eq:gamma-Dp_7-p}. 

To summarize, the derivation of the theory presented in this section consists of two steps:
\begin{align}
 \text{1st step:}\quad&\bigl(G_{ij},\,\phi,\,C^{(7-p)}\bigr)
 \overset{\text{exotic duality}}\longrightarrow 
 \bigl(\widetilde{G}_{ij},\,\widetilde{\bphi},\,\widetilde{C}^{(7-p)}\bigr) \,, 
\\
 \text{2nd step:}\quad&\bigl(\widetilde{G}_{ij},\,\widetilde{\bphi},\,\widetilde{C}^{(7-p)}\bigr) 
 \overset{\text{field redefinition}}{\longrightarrow} 
 \bigl(\tilde{g}_{ij},\,\tilde{\phi},\,\gamma^{i_1\cdots i_{7-p}}\bigr) \,. 
\end{align}
The first step is the exotic duality which is just a $U$-duality transformation that maps 
a non-geometric background to a standard geometric background. 
The second step is given by the field redefinition \eqref{eq:relations} 
which converts a $(7-p)$-form field into a $(7-p)$-vector, 
and after the redefinition, the action has the form \eqref{eq:gamma-action-p}. 

Although we do not know the general expression for the transformation rule under the exotic duality, 
for example, in the case of $p=5$ where the transformation is given by \eqref{eq:exotic-p=5}, 
we can obtain the following relation between 
$\bigl(\tilde{g}_{ij},\,\tilde{\phi},\,\gamma^{i_1\cdots i_{7-p}}\bigr)$ 
and $\bigl(G_{ij},\,\phi,\,C^{(7-p)}\bigr)$:
\begin{align}
\begin{split}
 (\tilde{g}_{ij})&= \frac{\abs{\det G_{\bfp\bfq}}^{1/2}}{\abs{\det F_{\bfp\bfq}}^{1/2}}\,\Exp{-2\phi}
 \bpm \Exp{-2\phi}\,G_{\bfa\bfb} & 0 \\ 0 & F_{\bfp\bfr}\,F_{\bfq\bfs}\,G^{\bfr\bfs}\epm
\\
 &= \frac{\abs{\det (\Exp{-\phi}G_{ij})}^{1/2}}{\abs{\det F_{ij}}^{1/2}}\,\Exp{2\phi}
 \bpm F_{\bfa k}\,F_{\bfb l}\,G^{kl} & 0 \\ 0 & F_{\bfp k}\,F_{\bfq l}\,G^{kl}\epm \,,
\\
 \Exp{2\tilde{\phi}}&= \frac{\abs{\det G_{\bfp\bfq}}}{\abs{\det F_{\bfp\bfq}}}\,\Exp{-6\phi} 
 = \frac{\abs{\det (\Exp{-\phi}G_{ij})}}{\abs{\det F_{ij}}}\,\Exp{2\phi} \,,
\\
 \gamma^{\bfp\bfq} &= (F^{-1})^{\bfp\bfr}\,(F^{-1})^{\bfq\bfs}\,C^{(2)}_{\bfr\bfs}
  = (F^{-1})^{\bfp i}\,(F^{-1})^{\bfq j}\,C^{(2)}_{ij}\,,
\end{split}
\end{align}
where we used the assumptions $G_{\bfa\bfp}=C^{(2)}_{\bfa i}=0$\,. 
This is precisely equal to the relation \eqref{eq:gamma-field-redefinition}, 
and the presentation given in this section serves as an alternative derivation of 
the theory \eqref{eq:gamma-action-reduced} with its generalization for general values of $p$ (with $p\neq 3$). 

\paragraph*{\underline{Duality rules for the new background fields}\\}

Here, we comment on the duality rules for the new background fields $\bigl(\tilde{g}_{ij},\,\tilde{\phi},\,\gamma^{i_1\cdots i_{7-p}}\bigr)$ 
under the $U$-duality transformations. 
By the construction of the theory presented in this section, 
it will be natural to define the transformation rule in the following manner. 

Let us begin with a configuration $\bigl(\tilde{g}_{ij},\,\tilde{\phi},\,\gamma^{i_1\cdots i_{7-p}}\bigr)$ 
which satisfies the equations of motion derived from the action \eqref{eq:gamma-action-p}. 
In order to perform a $U$-duality, we first transform these fields into $\bigl(\widetilde{G}_{ij},\,\widetilde{\bphi},\,\widetilde{C}^{(7-p)}\bigr)$ 
using the relation \eqref{eq:inv-relations}. 
Secondly, we use the standard transformation rules under the $U$-duality to obtain a new background 
$\bigl(\widetilde{G}'_{ij},\,\widetilde{\bphi}',\,\widetilde{C}'^{(7-p')},\cdots\bigr)$, 
where the ellipsis represents possible additional fields such as the $B$-field. 
Finally, using the relation \eqref{eq:relations},%%%%%%%%%%%%%%%%%%%%%%%%%%%%%%%%%%%%%%%
\footnote{Note that the relation \eqref{eq:relations} should be generalized if additional fields, such as the $B$-field, have non-vanishing values.} 
%%%%%%%%%%%%%%%%%%%%%%%%%%%%%%%%%%%%%%%%%%%%%%%%%%%%%%%%%%%%%%%%%%%%%%%%%%%%%%%%%%%%%%%%
we can obtain the $U$-dual background 
$\bigl(\tilde{g}'_{ij},\,\tilde{\phi}',\,\gamma'^{i_1\cdots i_{7-p'}},\cdots\bigr)$. 
At the same time, the original background fields $\bigl(G_{ij},\,\phi,\,C^{(7-p')}\bigr)$ should be transformed into $\bigl(G'_{ij},\,\phi',\,C'^{(7-p')}\bigr)$, 
which is the exotic dual of the background 
$\bigl(\widetilde{G}'_{ij},\,\widetilde{\bphi}',\,\widetilde{C}'^{(7-p')},\cdots\bigr)$. 

As an example, let us consider a simple configuration given by
\begin{align}
 \rmd \tilde{s}^2 = \tilde{g}_{\bfa\bfb}\,\rmd x^\bfa \rmd x^\bfb + \tilde{g}_{pq}\,\rmd x^p \rmd x^q 
  + \tilde{g}_{99}\,\rmd x_9^2 \,, \quad 
 \gamma^{(p+2)\cdots 8} = a\qquad (p,\,q=0,3\dotsc,8)\,.
\end{align}
In this case, if we take a $T$-duality in the $x^9$-direction, from the above procedure, 
we can straightforwardly obtain the following $T$-dual background:
\begin{align}
\begin{split}
 \rmd \tilde{s}'^2 &= \Bigl(\Exp{\frac{4\tilde{\phi}}{(p-3)(p-4)}+\frac{4\eta}{p-3}}\,\tilde{g}_{99}^{-\frac{2}{p-4}}\Bigr)\,\tilde{g}_{\bfa\bfb}\,\rmd x^\bfa \rmd x^\bfb + \tilde{g}_{pq}\,\rmd x^p \rmd x^q 
  + \Exp{-2\eta}\,\tilde{g}_{99}^{-1}\,\rmd x_9^2 \,, 
\\
 \Exp{2\tilde{\phi}'}&= \Exp{2\tilde{\phi}+2\eta}\,\tilde{g}_{99}^{-1} \,,\quad \gamma'^{(p+2)\cdots 9} = a\,.
\end{split}
\end{align}
We can easily confirm that this relation indeed maps the $\DD p_{7-p}(3\cdots(p+1)9,(p+2)\cdots 8)$ background \eqref{eq:gamma-Dp_7-p} 
to the $\DD (p-1)_{8-p}(3\cdots(p+1),(p+2)\cdots 9)$ background. 
In this way, we can show that the $\gamma$-fields follow the rule, 
$\gamma^{i_1\cdots i_p}\to \gamma^{i_1\cdots i_py}$ or $\gamma^{i_1\cdots i_py}\to\gamma^{i_1\cdots i_p}$, 
under the $T_y$-duality transformation, 
which is similar to the rule for the Ramond-Ramond fields. 

In the case of general $U$-duality transformation, 
since multiple Ramond-Ramond fields and the $B$-field have non-vanishing values, 
we can no more use the relation \eqref{eq:relations}. 
However, as we discuss in the next subsection, in the case of an $S$-duality transformation, 
we can obtain a relation similar to \eqref{eq:relations} and derive the $\beta$-supergravity. 

\subsection{Effective actions for $Q$-fluxes}
\label{sec:NS-NS}

We here present another derivation of the (simplified) action \eqref{eq:beta-action-reduced} 
for the $\beta$-supergravity from the action \eqref{eq:gamma-action-p}. 
We further obtain an action for a non-geometric flux which has an exotic string 
(i.e.~$1^6_4$-brane) as the magnetic source. 

\paragraph*{\underline{Alternative derivation of the $\beta$-supergravity}\\}

Let us consider an $S$-duality transformation given by
\begin{align}
\begin{alignedat}{3}
 \widetilde{G}_{ij} &\to \Exp{-\widetilde{\bphi}}\, \widetilde{G}_{ij}\,,\quad&
 \widetilde{\bphi} &\to - \widetilde{\bphi}\,,\quad& \widetilde{C}^{(2)}_{ij}&\to -\widetilde{B}_{ij} \,,
\\
 \tilde{g}_{ij} &\to \Exp{-\tilde{\phi}}\, \tilde{g}_{ij}\,, \quad&
 \tilde{\phi}&\to -\tilde{\phi}\,,\quad& \gamma^{ij}&\to -\beta^{ij}\,,\quad \eta\to -\eta\,.
\end{alignedat}
\end{align}
Under the redefinition of fields, the action \eqref{eq:gamma-action-p} and the relation \eqref{eq:relations} with $p=5$ become
\begin{align}
 S &= \frac{1}{2\kappa_{10}^2}\,
    \int \biggl[\Exp{-2\tilde{\phi}}\, \bigl(\tilde{*}\,\tilde{R} + 4\,\rmd \tilde{\phi}\wedge \tilde{*}\,\rmd \tilde{\phi}\bigr)
 - \frac{1}{4}\,\Exp{-2\tilde{\phi}}\,\bQ^{(1)\,ij}\wedge \tilde{*}\,\bQ^{(1)}_{ij}\biggr]\,,
\\
\begin{split}
 (\tilde{g}_{ij}) 
 &= \begin{pmatrix} \widetilde{G}_{\bfa\bfb}&0\\
 0& \widetilde{G}^{\bfp\bfq}
\end{pmatrix} \,,
\quad
 \Exp{2\tilde{\phi}} = \Exp{-2\widetilde{\bphi}+2\eta} \,,
\quad
 \beta^{ij}=\widetilde{B}_{ij}\,,
\\
 \Exp{\eta}&= \Bigl(\Exp{-2\widetilde{\bphi}}\sqrt{\abs{\det \widetilde{G}_{\bfp\bfq}}}\Bigr)^{-1} \,.
\label{eq:beta-tilde-widetilde}
\end{split}
\end{align}
Further, under the $S$-duality, the transformation rule for the exotic duality given in \eqref{eq:exotic-p=5} becomes
\begin{align}
 \bigl(G_{ij},\,&\phi,\,B^{(2)}\bigr)
 \overset{\text{exotic duality}}\longrightarrow 
 \bigl(\widetilde{G}_{ij},\,\widetilde{\bphi},\,\widetilde{B}^{(2)}\bigr) \,,
\\
\begin{split}
 (\tilde{G}_{ij}) 
 &= \begin{pmatrix} 
 G_{\bfa\bfb}&0 \\ 0& \bigl(E^{-1}\bigr)^{\bfp i}\,\bigl(E^{-1}\bigr)^{\bfq j}\,G_{ij}
\end{pmatrix}\,, 
\\
 \tilde{B}_{\bfp\bfq} &\equiv \bigl(E^{-1}\bigr)^{\bfp i}\,\bigl(E^{-1}\bigr)^{\bfq j}\,B_{ij}\,,\quad 
 \Exp{-2\tilde{\bphi}}\sqrt{\abs{\det\tilde{G}_{\bfp\bfq}}}
 \equiv \Exp{-2\phi}\sqrt{\abs{\det G_{\bfp\bfq}}}\,.
\end{split} 
\end{align}
Substituting this relation into \eqref{eq:beta-tilde-widetilde}, 
we can correctly reproduce the transformation rule \eqref{eq:beta-fields} 
between $\bigl(G_{ij},\,\phi,\,B^{(2)}\bigr)$ and $\bigl(\tilde{g}_{ij},\,\tilde{\phi},\,\beta^{ij}\bigr)$. 
This gives another derivation of the action \eqref{eq:beta-action-reduced} in the case $G_{\bfa\bfp}=B_{\bfa i}=0$ is satisfied. 

\paragraph*{\underline{Supergravity action for the exotic string $1^6_4$}\\}

Now, let us consider applying an $S$-duality transformation to the action \eqref{eq:gamma-action-p} with $p=1$. 
By using the $S$-duality transformation rule,
\begin{align}
\begin{alignedat}{3}
 G_{ij} &\to \Exp{-\phi}\, G_{ij}\,,\quad&
 \phi &\to - \phi\,,\quad& C^{(6)}_{i_1\cdots i_6}&\to B_{i_1\cdots i_6} \,,
\\
 \tilde{g}_{ij} &\to \Exp{-\tilde{\phi}}\, \tilde{g}_{ij}\,, \quad&
 \tilde{\phi}&\to -\tilde{\phi}\,,\quad& \gamma^{i_1\cdots i_6}&\to \beta^{i_1\cdots i_6}\,,\quad \eta\to -\eta\,,
\end{alignedat}
\end{align}
the action \eqref{eq:gamma-action-p} and the relation \eqref{eq:relations} with $p=1$ become
\begin{align}
 &S= \frac{1}{2\kappa_{10}^2}\,
    \int \biggl[\Exp{-2\tilde{\phi}}\, \bigl(\tilde{*} \tilde{R} + 4\,\rmd \tilde{\phi}\wedge \tilde{*}\,\rmd \tilde{\phi}\bigr)
 - \frac{1}{2\cdot 6!}\,\Exp{-6\tilde{\phi}}\bQ^{(1)\,i_1\cdots i_{6}}\wedge \tilde{*}\,\bQ^{(1)}_{i_1\cdots i_{6}}\biggr] \,,
\label{eq:exotic-string-action}
\\
\begin{split}
 &(\tilde{g}_{ij}) 
 = \begin{pmatrix} \Exp{-4\widetilde{\bphi}+6\eta} \widetilde{G}_{\bfa\bfb}&0\\
 0&\Exp{2\eta} \widetilde{G}^{\bfp\bfq}
\end{pmatrix} \,, \quad 
 \Exp{2\tilde{\phi}} = \Exp{-2\widetilde{\bphi}+6\eta} \,,
\\
 &\beta^{\bfs_1\cdots \bfs_6} =\widetilde{B}^{(6)}_{\bfs_1\cdots \bfs_6}\,, \quad
 \Exp{\eta} = \Exp{-2\widetilde{\bphi}}\sqrt{\abs{\det \widetilde{G}_{\bfp\bfq}}} \,.
\label{eq:exotic-string-relation}
\end{split}
\end{align}
In this case, it is difficult to find a transformation rule for the exotic duality
\begin{align}
 \bigl(G_{ij},\,\phi,\,B^{(6)}\bigr)
 \overset{\text{exotic duality}}\longrightarrow 
 \bigl(\widetilde{G}_{ij},\,\widetilde{\bphi},\,\widetilde{B}^{(6)}\bigr) \,.
\end{align}
However, as a special case, if we consider the $1^6_4(3,456789)$ background, 
the fields with tilde $\bigl(\widetilde{G}_{ij},\,\widetilde{\bphi},\,\widetilde{B}^{(6)}\bigr)$ 
should have the same form with those in the $\FF1(3)$ background. 
Then, from the relation \eqref{eq:exotic-string-relation}, 
we can obtain the following solution of the theory \eqref{eq:exotic-string-action}, 
which corresponds to the background of the $1^6_4(3,456789)$-brane:
\begin{align}
\begin{split}
 \underline{\text{ $1^6_4$ :}}\quad 
 &\rmd \tilde{s}^2 = \rho_2\,\bigl(\rho_2\,\abs{f}^2\,\rmd z\,\rmd \bar{z} +\rmd x^2_{03}\bigr) + \rmd x_{4\cdots 9}^2 \,, \quad
 \Exp{2\tilde{\phi}} = \rho_2 \,, 
\\
 &\beta^{456789}= \rho_1 \,, \quad \beta_{456789}^{(8)} = - \rho_2^{-1}\,\rmd t\wedge \rmd x^3\wedge \cdots \wedge \rmd x^9\,.
\end{split}
\end{align}
We can also find a Euclidean solution which corresponds to the background of an instanton that electrically couples to $\beta^{i_1\cdots i_6}$:
\begin{align}
\begin{split}
 &\rmd \tilde{s}^2 = \rho_2^{-1}\, \bigl(\abs{f}^2\,\rmd z\,\rmd \bar{z}+\rmd\tau^2+\rmd x^2_{n_1n_2}\bigr) + \rmd x^2_{m_1\cdots m_6} \,, 
\quad
 \Exp{2\tilde{\phi}} = \rho_2^{-1} \,, 
\\
 &\beta^{(8)}_{m_1\cdots m_6} = -\rho_1\,\rmd \tau \wedge \rmd x^3\wedge \cdots \wedge \rmd x^9 \,, \quad 
 \beta^{m_1\cdots m_6} = \rho_2^{-1} \,.
\end{split}
\label{eq:beta6-instantons}
\end{align}

\subsection{Seven branes and instantons}
\label{sec:p=7}

In the case of $p=7$, the exotic duality is the same as the $S$-duality:
\begin{align}
 \widetilde{G}_{ij}=\abs{\tau}\,G_{ij}\,,\quad \Exp{2\widetilde{\bphi}}= \abs{\tau}^4\,\Exp{2\phi} \,,\quad \widetilde{C}^{(0)}=-\frac{C^{(0)}}{\abs{\tau}^2} \,.
\end{align}
In this case, from the relation \eqref{eq:relations} with $p=7$, 
we can write down an explicit relation between $\bigl(G_{ij},\,\phi,\,C^{(0)}\bigr)$ and 
$\bigl(\tilde{g}_{ij},\,\tilde{\phi},\,\gamma\bigr)$ as in the case of $p=5$\,:
\begin{align}
\begin{split}
 (\tilde{g}_{ij}) 
 &= \begin{pmatrix} \abs{\tau}^{-2}\,\Exp{-\phi}\, \tilde{G}_{\bfa\bfb}&0\\ 0&\Exp{\eta} \tilde{G}^{\bfp\bfq}
\end{pmatrix} \,,
\quad
 \Exp{2\tilde{\phi}} = \frac{\Exp{-2\phi}}{\abs{\tau}^4} \,,\quad 
 \gamma = \widetilde{C}^{(0)} = -\frac{C^{(0)}}{\abs{\tau}^2} \,,
\\
 \Exp{2\eta}&= \Exp{-2\bphi}\sqrt{\abs{\det G_{\bfp\bfq}}} \,.
\end{split}
\end{align}
From this relation, we can obtain the expression for $\bP^{(9)}$ in terms of the original background fields $\bigl(G_{ij},\,\phi,\,C^{(0)}\bigr)$:
\begin{align}
 \bP^{(9)} 
 &= \Exp{-4\tilde{\phi}} \,\tilde{*}\,\rmd \gamma 
  = \Exp{-4\tilde{\phi}}\,\frac{\sqrt{\abs{\tilde{g}}}}{9!}\,\tilde{g}^{\bfa\bfb}\,\partial_\bfa \gamma\,\epsilon_{\bfb i_1\cdots i_9}\,\rmd x^{i_1}\wedge \cdots\wedge \rmd x^{i_9}
\nn\\
 &= \abs{\tau}^4\,\frac{\sqrt{\abs{\tilde{G}}}}{9!}\,\tilde{G}^{\bfa\bfb}\,\partial_\bfa \gamma \,\epsilon_{\bfb i_1\cdots i_9}\,\rmd x^{i_1}\wedge \cdots\wedge \rmd x^{i_9}
  = \abs{\tau}^4 *\,\rmd \biggl(-\frac{C^{(0)}}{\abs{\tau}^2}\biggr) 
\nn\\
 &= -C^{(0)}\,\rmd B^{(8)} + \abs{\tau}^2\, \rmd C^{(8)}\,,
\label{eq:P9-expression}
\end{align}
where we defined \cite{Eyras:1999at}
\begin{align}
 \rmd B^{(8)}\equiv *\,\rmd \abs{\tau}^2 = \Exp{2\phi}*_{\EE} \rmd \abs{\tau}^2\,,\quad 
 \rmd C^{(8)}\equiv *\,\rmd C^{(0)} =\Exp{2\phi}*_{\EE}\rmd C^{(0)}\,.
\end{align}
Then, the equation of motion \eqref{eq:EOM3} can be written as
\begin{align}
 \rmd \bP^{(9)} = \rmd \bigl(-C^{(0)}\,\rmd B^{(8)} + \abs{\tau}^2\, \rmd C^{(8)}\bigr)=0\,.
\end{align}
In fact, this equation of motion can be derived, in the original theory, 
as a conservation law for the Noether current associated with the $\SL(2)$ symmetry \cite{Meessen:1998qm}. 
Conversely, in the theory \eqref{eq:gamma-action-p}, the equation of motion, $\rmd F^{(9)}=\rmd^2 C^{(8)}=0$, 
of the original theory 
will appear as a conservation law for the Noether current. 

As it was shown in \cite{Eyras:1999at}, the dual potential $\gamma^{(8)}$ (or $\widetilde{C}^{(8)}$ given in \cite{Eyras:1999at}) 
is the field that electrically couples to an NS7-brane. 
This can also be understood from the fact that a D7-brane couples to $C^{(8)}$ 
and the 8-forms transform under an $S$-duality (or the exotic duality) as follows \cite{Eyras:1999at}:
\begin{align}
 C^{(8)} \to -\gamma^{(8)}\,,\quad \gamma^{(8)}\to -C^{(8)}\,,\quad B^{(8)}\to -B^{(8)} \,.
\end{align}

Further, as it is well-known, a D-instanton, or a $\DD(-1)$-brane, electrically couples to the Ramond-Ramond 0-form, $C^{(0)}$. 
The ``mass,'' i.e., the on-shell value of the Euclidean action is calculated in \cite{Gibbons:1995vg} 
and is proportional to $\gs^{-1}$, like the tension of the D$p$-branes. 
If we perform an $S$-duality, the D-instanton is mapped to another instanton 
which couples to the field $\gamma=\widetilde{C}^{(0)}$ 
and will have the ``mass'' proportional to $\gs$, since the $S$-duality maps $\gs\to 1/\gs$. 
Indeed, in the Einstein frame, the action for $\gamma$ becomes
\begin{align}
 S = \frac{1}{2\kappa_{10}^2}\,\int \sqrt{\abs{\tilde{g}_{\EE}}}\, \Bigl(\tilde{*}_\EE \tilde{R}_{\EE} -\frac{1}{2}\,\rmd\tilde{\phi}\wedge *_{\EE}\rmd\tilde{\phi} 
      - \frac{1}{2}\,\Exp{-2\tilde{\phi}}\, \rmd \gamma\wedge *_{\EE}\rmd\gamma \Bigr) \,,
\end{align}
and using the result (6.6) of \cite{Bergshoeff:2004fq}, we can find that the value of the on-shell action is proportional to $\gs$. 
In the following, we will denote the instanton by $\II_{1}$\,. 
Note that this instanton is a special case of the $(p,q)$-instanton, 
which is a member of the \emph{$Q$-instantons} discovered in \cite{Bergshoeff:2008qq}. 
By performing $T$-dualities in the $x^\bfp$-directions, 
$\gamma$ will be mapped to a $(7-p)$-vector $\gamma^{i_1\cdots i_{7-p}}$, 
and we will obtain an instanton, to be called $\II^{7-p}_{1}$, which electrically couples to $\gamma^{i_1\cdots i_{7-p}}$. 
Recalling the transformation rule for the fundamental constants under the action of $T$-duality, $\gs\to \gs\, (\ls/R_{i_n})$ and $\ls\to \ls$, 
we expect that the ``mass'' of $\II^{7-p}_{1}$ will be proportional to $\gs\,(\ls^{7-p}/R_{\bfp_1}\cdots R_{\bfp_{7-p}})$ 
(see section 7 of \cite{Bergshoeff:2011se} for a discussion on the existence of such objects). 
The family of instantons, $\II^{7-p}_{1}$, can be thought of as generalizations of the pp-wave (whose mass is proportional to $(\ls/R_i)$), 
much like the exotic branes can be thought of as generalizations of the Kaluza-Klein monopole. 
The instanton $\II^{7-p}_{1}$ can exist only when there are $(7-p)$ compact isometry directions. 
Since it is the electric source of $\bP^{(1)\,i_1\cdots i_{7-p}}$, 
the corresponding background should be given by \eqref{eq:gamma-instantons}. 

Further, in the type IIB theory, by performing an $S$-duality, 
the background fields $\gamma^{ij}$ and $\gamma^{i_1\cdots i_6}$ are transformed into $\beta^{ij}$ and $\beta^{i_1\cdots i_6}$\,. 
At the same time, the instantons $\II^2_{1}$ and $\II^6_{1}$ will be mapped to other instantons, to be called $\II^2_0$ and $\II^6_{2}$, 
whose ``mass'' will be proportional to $(\ls^2/R_{\bfp_1} R_{\bfp_2})$ and $\gs^2\,(\ls^{6}/R_{\bfp_1}\cdots R_{\bfp_6})$, respectively. 
These instantons are also predicted in \cite{Bergshoeff:2011se}. 
The corresponding background solutions will be given by \eqref{eq:beta-2-instanton} and \eqref{eq:beta6-instantons}.

\subsection{The mixed-symmetry tensors}

So far, we argued that the mixed-symmetry tensors 
$\beta^{(8)}_{ij}$, $\beta^{(8)}_{i_1\cdots i_6}$, and $\gamma^{(8)}_{i_1\cdots i_{7-p}}$ 
electrically couple to exotic branes. 
Here, we provide expressions for these mixed-symmetry tensors in terms of the original background fields, 
like the relation \eqref{eq:P9-expression}. 

For convenience, we denote the form fields $B^{(2)}$, $B^{(6)}$, and $C^{(7-p)}$ collectively as $\sfA^{(7-p)}$, 
and the mixed-symmetry tensors as $\cA^{i_1\cdots i_{7-p}}$ and $\cA^{(8)}_{i_1\cdots i_{7-p}}$\,. 
We again use the following tilde notation for the background fields which are related by the exotic duality:
\begin{align}
 \bigl(G_{ij},\,\phi,\,\sfA^{(7-p)}\bigr)
  \overset{\text{exotic duality}}\longrightarrow 
 \bigl(\widetilde{G}_{ij},\,\widetilde{\bphi},\,\widetilde{\sfA}^{(7-p)}\bigr) \,.
\end{align}
In the following, we derive an expression for the mixed tensor $\cA^{(8)}_{i_1\cdots i_{7-p}}$ 
in terms of the background fields on the right-hand sides; 
$\bigl(\widetilde{G}_{ij},\,\widetilde{\bphi},\,\widetilde{\sfA}^{(7-p)}\bigr)$. 

We define the dual potential $\widetilde{\sfA}^{(p+1)}$ by
\begin{align}
 \rmd\widetilde{\sfA}^{(p+1)}\equiv s_p\,\Exp{2\,(\alpha_s+1)\,\widetilde{\bphi}}\,\tilde{*}\,\rmd \widetilde{\sfA}^{(7-p)}\,,
\end{align}
where $s_p\equiv (-1)^{\frac{(p+2)(p+1)}{2}+\alpha_s+1}$ and 
$\alpha_s=\{-1,\,-2,\,0\}$ for $\widetilde{\sfA}^{(p+1)}=\{\widetilde{C}^{(p+1)},\,\widetilde{B}^{(2)},\,\widetilde{B}^{(6)}\}$, 
and $\tilde{*}$ is the Hodge star operator associated with $\widetilde{G}_{ij}$\,. 
Then, using the identification 
$\cA^{i_1\cdots i_{7-p}}\equiv\{\gamma^{i_1\cdots i_{7-p}},\,\beta^{i_1i_2},\,\beta^{i_1\cdots i_6}\}
 = \widetilde{\sfA}^{(7-p)}_{i_1\cdots i_{7-p}}$ and the definition
\begin{align}
 \rmd \cA^{(8)}_{i_1\cdots i_{7-p}}\equiv 
 \Exp{2\,(\alpha+1)\,\tilde{\phi}}\,\tilde{g}_{i_1j_1}\cdots\tilde{g}_{i_{7-p}j_{7-p}}\,\tilde{*}\,\rmd \cA^{j_1\cdots j_{7-p}}\quad
 (\alpha\equiv -\alpha_s-4)\,,
\end{align}
we obtain
\begin{align}
 \partial_\bfa \widetilde{\sfA}^{(p+1)}_{\bfs_1\cdots \bfs_{p+1}}
 &= s_p\,\Exp{2\,(\alpha_s+1)\,\widetilde{\bphi}}\,\frac{\sqrt{\abs{\widetilde{G}}}}{(7-p)!}\,
    \widetilde{G}^{\bfc\bfb}\,\widetilde{G}^{\bfq_1\bfr_1}\cdots\widetilde{G}^{\bfq_{7-p}\bfr_{7-p}}\,
    \epsilon_{\bfc\bfq_1\cdots \bfq_{7-p}\bfa\bfs_1\cdots \bfs_{p+1}}\,
    \partial_\bfb \widetilde{\sfA}^{(7-p)}_{\bfr_1\cdots \bfr_{7-p}}
\nn\\
 &= s_p\,\Exp{2\,(\alpha+1)\,\tilde{\phi}}\,\frac{\sqrt{\abs{\tilde{g}}}}{(7-p)!}\,\tilde{g}^{\bfc\bfb}\,
    \tilde{g}_{\bfq_1\bfr_1}\cdots\tilde{g}_{\bfq_{7-p}\bfr_{7-p}}\,\epsilon_{\bfc\bfq_1\cdots \bfq_{7-p}\bfa\bfs_1\cdots \bfs_{p+1}}\,
    \partial_\bfb \cA^{\bfr_1\cdots \bfr_{7-p}} 
\nn\\
 &= s_p\,(-1)^{(p+2)(7-p)}\,\partial_\bfa \cA^{(8)}_{\bfs_1\cdots \bfs_{p+1}\bfq_1\cdots \bfq_{7-p},\bfq_1\cdots \bfq_{7-p}} \,,
\end{align}
where the summation over the indices $\bfq_n$ is assumed in the second line but not in the third line, 
and we used
\begin{align}
 \widetilde{G}_{\bfa\bfb}\propto \tilde{g}_{\bfa\bfb}\,,\quad 
 \widetilde{G}^{\bfp\bfq}=\Exp{-(\alpha+2)\,\eta}\,\tilde{g}_{\bfp\bfq}\,,\quad 
 \Exp{\eta}=\Bigl(\Exp{-2\widetilde{\bphi}}\sqrt{\abs{\det \widetilde{G}_{\bfp\bfq}}}\Bigr)^{\frac{2}{p-3-4(\alpha+3)}} \,,
\end{align}
in the second equality. 
Namely, we obtain
\begin{align}
\begin{split}
 &\partial_\bfa \cA^{(8)}_{\bfs_1\cdots \bfs_{p+1}\bfq_1\cdots \bfq_{7-p},\bfq_1\cdots \bfq_{7-p}}
  = s_p\,(-1)^{(p+2)(7-p)}\,\partial_\bfa \sfA^{(p+1)}_{\bfs_1\cdots \bfs_{p+1}}
\\
 &=(-1)^{(p+2)(7-p)}\,
   \frac{\Exp{-2(\alpha+3)\tilde{\bphi}}}{(7-p)!}\,\varepsilon^{\bfb\bfr_1\cdots \bfr_{7-p}}{}_{\bfa\bfs_1\cdots \bfs_{p+1}}\,
   \partial_\bfb \sfA^{(7-p)}_{\bfr_1\cdots \bfr_{7-p}}\quad \bigl(\varepsilon_{i_0\cdots i_9}\equiv \sqrt{\abs{\tilde{g}}}\,\epsilon_{i_0\cdots i_9}\bigr)\,,
\end{split}
\end{align}
which reproduces the proposed relations (30)--(32) of \cite{Bergshoeff:2011se}. 
If we further use the relation between $\bigl(G_{ij},\,\phi,\,\sfA^{(7-p)}\bigr)$ and $\bigl(\widetilde{G}_{ij},\,\widetilde{\bphi},\,\widetilde{\sfA}^{(7-p)}\bigr)$, 
we can also obtain the relation between $\Co\cA^{(8)}_{\bfq_1\cdots \bfq_{7-p}}$ 
and $\bigl(G_{ij},\,\phi,\,\sfA^{(7-p)}\bigr)$. 

\subsection{Summary of the results}

In this section, we have presented various actions with the following form:
\begin{align}
 S\bigl[\tilde{g}_{ij},\,\tilde{\phi},\,\cA^{i_1\cdots i_{7-p}}\bigr] 
 &= \frac{1}{2\kappa_{10}^2}\,
    \int \biggl[\Exp{-2\phi}\,\bigl(\tilde{*}\,\tilde{R} + 4\,\rmd \phi\wedge \tilde{*}\,\rmd \phi\bigr)
\nn\\
 &\quad\qquad\qquad 
 - \frac{\Exp{2\,(\alpha+1)\,\tilde{\phi}}}{2\,(7-p)!}\,\tilde{g}_{i_1j_1}\cdots \tilde{g}_{i_{7-p}j_{7-p}}\,
   \cQ^{(1)\,i_1\cdots i_{7-p}}\wedge \tilde{*}\, \cQ^{(1)\,j_1\cdots j_{7-p}}\biggr]\,,
\label{eq:action-general}
\end{align}
where $\cQ^{(1)\,i_1\cdots i_{7-p}} \equiv \rmd\cA^{i_1\cdots i_{7-p}}$ is a non-geometric flux 
of which an exotic brane acts as the magnetic source,
and $\alpha$ is an integer given in Table \ref{tab:exotic}. 
\begin{table}[t]
\centering
 \begin{tabular}{|c||c|c|c|}\hline
 $\cA^{i_1\cdots i_{7-p}}$ & $\beta^{ij}$ & $\gamma^{i_1\cdots i_{7-p}}$ & $\beta^{i_1\cdots i_6}$ \\\hline
 $\cQ^{(1)\,i_1\cdots i_{7-p}}$ & \quad$\bQ^{(1)\,ij}$\quad & \quad$\bP^{(1)\,i_1\cdots i_{7-p}}$\quad & \quad$\bQ^{(1)\,i_1\cdots i_6}$\quad \\\hline
 $\alpha$ & $-2$ & $-3$ & $-4$ \\\hline
 magnetic source $\bigl(p^{7-p}_{-\alpha}\bigr)$ & $5^2_2$ & $p^{7-p}_3$ & $1^6_4$ \\\hline
 $\tilde{\alpha}\equiv -\alpha-2$ & $0$ & $1$ & $2$ \\\hline
 electric source $\bigl(\II^{7-p}_{\tilde{\alpha}}\bigr)$ & $\II^2_0$ & $\II^{7-p}_{1}$ & $\II^6_{2}$ \\\hline
 \end{tabular}
\caption{A list of non-geometric fluxes and their magnetic/electric sources.}
\label{tab:exotic}
\end{table}
The equations of motion are given by
\begin{align}
 &\tilde{R}+4\bigl(\tilde{\nabla}^i\partial_i\tilde{\phi}-\tilde{g}^{ij}\,\partial_i\tilde{\phi}\,\partial_j\tilde{\phi}\bigr)
 +\frac{(\alpha+1)\,\Exp{2\,(\alpha+2)\,\tilde{\phi}}}{2\,(7-p)!}\,\cQ_i{}^{j_1\cdots j_{7-p}}\,\cQ^i{}_{j_1\cdots j_{7-p}} =0 \,,
\\
 &\tilde{R}_{ij} +2\tilde{\nabla}_i\partial_j\tilde{\phi}-\frac{\Exp{2\,(\alpha+2)\,\tilde{\phi}}}{2\,(7-p)!}\,
 \Bigl(\cQ_i{}^{k_1\cdots k_{7-p}}\,\cQ_j{}_{k_1\cdots k_{7-p}} -(7-p)\,\cQ_{k_1}{}^{k_2\cdots k_{7-p}}\,\cQ^{k_1}{}_{jk_2\cdots k_{7-p}}
\nn\\
 &\qquad\qquad\qquad\qquad\qquad\qquad -\frac{\alpha+2}{2}\,\cQ_k{}^{l_1\cdots l_{7-p}}\,\cQ^k{}_{l_1\cdots l_{7-p}} \,\tilde{g}_{ij} \Bigr) =0 \,,
\\
 &\rmd \cQ^{(9)}_{i_1\cdots i_{7-p}}=0\,,\quad 
 \cQ^{(9)}_{i_1\cdots i_{7-p}} \equiv 
 \Exp{2\,(\alpha+1)\,\tilde{\phi}}\, \tilde{g}_{i_1j_1}\cdots\tilde{g}_{i_{7-p}j_{7-p}}\, \tilde{*}\, \cQ^{(1)\,j_1\cdots j_{7-p}} 
 \equiv \rmd\cA^{(8)}_{i_1\cdots i_{7-p}}\,.
\end{align}
If we regard the dual potential $\cA^{(8)}_{i_1\cdots i_{7-p}}$ as a fundamental field, 
the dual action is given by
\begin{align}
\begin{split}
 S\bigl[\tilde{g}_{ij},\,\tilde{\phi},\,\cA^{(8)}_{i_1\cdots i_{7-p}}\bigr] 
 &= \frac{1}{2\kappa_{10}^2}\,
    \int \biggl[\Exp{-2\phi}\,\bigl(\tilde{*}\,\tilde{R} + 4\,\rmd \phi\wedge \tilde{*}\,\rmd \phi\bigr)
\\
 &\quad\qquad\qquad 
 - \frac{\Exp{2\,(\tilde{\alpha}+1)\,\tilde{\phi}}}{2\,(7-p)!}\,\tilde{g}^{i_1j_1}\cdots \tilde{g}^{i_{7-p}j_{7-p}}\,
   \rmd \cA^{(8)}_{i_1\cdots i_{7-p}}\wedge \tilde{*}\, \rmd \cA^{(8)}_{j_1\cdots j_{7-p}}\biggr] \,,
\end{split}
\end{align}
where we defined $\tilde{\alpha}\equiv -\alpha - 2$\,. 
We can add the Wess-Zumino term of the exotic $p^{7-p}_{-\alpha}$-brane 
extending in the $x^{\bfr_1},\cdots,x^{\bfr_p}$-directions and 
smeared over the $x^{\bfs_1},\cdots,x^{\bfs_{7-p}}$-directions:
\begin{align}
 S_{\WZ} &= -\mu_{p^{7-p}_{-\alpha}}\,\sum_{\bfs_1,\cdots,\bfs_{7-p}}\int_{\cM_{p+1}\times T^{7-p}_{\bfs_1\cdots \bfs_{7-p}}}\frac{n^{\bfs_1\cdots\bfs_{7-p}}}{(7-p)!}\,\Co\cA_{\bfs_1\cdots \bfs_{7-p}}^{(8)}\wedge \frac{\rmd x^{\bfs_1}\wedge\cdots\wedge\rmd x^{\bfs_{7-p}}}{(2\pi R_{\bfs_1})\cdots (2\pi R_{\bfs_{7-p}})}
\nn\\
 &= -\mu_{p^{7-p}_{-\alpha}}\,\int\frac{1}{(7-p)!}\, \cA_{\bfs_1\cdots \bfs_{7-p}}^{(8)}\wedge \bdelta^{\bfs_1\cdots \bfs_{7-p}}(x-X(\xi)) \,.
\end{align}
Then, taking variation, we obtain the following Bianchi identity as the equation of motion:
\begin{align}
 \rmd^2 \cA^{\bfs_1\cdots \bfs_{7-p}} 
 = 2\kappa_{10}^2\,\mu_{p^{7-p}_{-\alpha}}\,\frac{n^{\bfs_1\cdots \bfs_{7-p}}}{(2\pi R_{\bfs_1})\cdots (2\pi R_{\bfs_{7-p}})}\,
   \delta^2(x-X(\xi))\,\rmd x^1\wedge\rmd x^2 \,. 
\end{align}
If we choose $n^{\bfs_1\cdots \bfs_{7-p}}=1$ and integrate the equation, we obtain
\begin{align}
 \sigma = \int\rmd^2 \cA^{\bfs_1\cdots \bfs_{7-p}} 
 = \frac{2\kappa_{10}^2\,\mu_{p^{7-p}_{-\alpha}}}{(2\pi R_{\bfs_1})\cdots (2\pi R_{\bfs_{7-p}})} \,,
\end{align}
where we used $\cA^{\bfs_1\cdots \bfs_{7-p}}=\rho_1$\,. 
From this relation and the value of $\sigma$ given in Table \ref{tab:sigma}, 
we can confirm that $\mu_{p^{7-p}_{-\alpha}}$ is indeed equal to the tension of the exotic brane:
\begin{align}
 \mu_{p^{7-p}_{-\alpha}} = \frac{\sigma\,(2\pi R_{\bfs_1})\cdots (2\pi R_{\bfs_{7-p}})}{(2\pi\ls)^7\,\ls\,\gs^2} 
     = \frac{M_{p^{7-p}_{-\alpha}}}{(2\pi R_{\bfr_1})\cdots (2\pi R_{\bfr_{p+1}})}\,,
\end{align}
where we used $2\kappa_{10}^2=(2\pi\ls)^7\,\ls\,\gs^2$\,. 

\section{Summary and discussions}
\label{sec:summary}

In this paper, we have presented (truncated) effective actions for various polyvectors 
whose magnetic sources can be identified with the exotic branes. 
Requiring the existence of compact isometry directions, 
which are required for the existence of exotic branes, 
we showed that the effective actions can be derived 
from the standard (truncated) supergravity actions. 
In each theory, we found two solutions with either the magnetic or the electric charge associated with the non-geometric flux. 
The former solution corresponds to an exotic-brane background 
while the latter corresponds to a new instanton background. 
By taking account of the $U$-duality symmetry of the string theory, 
all defect branes including the standard branes and exotic branes should be treated as equals. 
However, in the standard formulation of the supergravity, 
only the backgrounds of the standard branes are well described globally. 
Contrarily, in the reformulation of supergravity presented in this paper, 
the background fields of the exotic branes are globally defined 
while those of the standard branes are not single-valued. 
In this sense, the effective theory considered in this paper is complementary to the standard supergravity. 

Our reformulation is still not complete and we should investigate a further generalization 
so as to allow for general backgrounds with multiple non-geometric fluxes. 
Such generalization is necessary if we consider, for example, 
the background representing a bound state of $p$ $5^2_2$-branes and $q$ $5^2_3$-branes, 
which is the exotic dual of the $(p,q)$ five-brane (i.e., a bound state of $p$ NS5-branes and $q$ D5-branes). 
In addition, there is another direction of generalization. 
The action \eqref{eq:action-general} presented in this paper can be applied only for defect backgrounds 
with isometries in the $03\cdots 9$-directions. 
In the absence of these isometries, the action should be modified, like the action of the $\beta$-supergravity \eqref{eq:beta-action} 
or its Ramond-Ramond counterpart \eqref{eq:gamma-action}. 
Since the equations of motion derived from the simple action \eqref{eq:action-general} 
do not coincide with those derived from the complete action 
(i.e., the action \eqref{eq:beta-action} or \eqref{eq:gamma-action}) 
even though we impose the assumptions \eqref{eq:assumptions},%%%%%%%%%%%%%%
\footnote{%%%%%%%%%%%%%%%%%%%%%%%%%%%%%%%%%%%%%%%%%%%%%%%%%%%%%%%%%%%%%
We would like to thank David Andriot for pointing out this issue. 
In the case of the $\beta$-supergravity, 
if we assume that any derivative $\partial_i$ contracted with 
$\beta^{ij}$ vanishes, the dilaton equation of motion and 
the Einstein equations (see (1.28) and (1.29) of \cite{Andriot:2013xca}) 
coincide with those derived from the simple action \eqref{eq:beta-action} 
(note that $\widecheck{\cR}^{ij}= (1/4)\,\bigl(Q^{ikl}\,Q^{j}{}_{kl}-2\,Q^{kil}\,Q_k{}^j{}_l\bigr)$ and $\widecheck{\cR}=-(1/4)\,Q^{ijk}\,Q_{ijk}$ 
can be shown by using the simplifying assumption). 
However, the equation of motion for the $\beta$-field (see (1.30) of \cite{Andriot:2013xca}) 
does not coincide with \eqref{eq:beta-Bianchi}, 
although the difference disappears in the $5^2_2$ background. 
} 
%%%%%%%%%%%%%%%%%%%%%%%%%%%%%%%%%%%%%%%%%%%%%%%%%%%%%%%%%%%%%%%%%%%%%%%
it will be important to find the complete actions for any value of $p$ 
and check whether the exotic-brane backgrounds indeed satisfy 
the equations of motion derived from the complete actions.%$%%
\footnote{In the case of the $5^2_2$-brane (or the $Q$-brane), the background is shown to satisfy the equations of motion derived from the complete action 
(see Appendix D.1 of \cite{Andriot:2014uda}).} 
It will be also interesting to describe various non-geometric backgrounds with less isometries, 
such as the background of the ``NS5-brane localized in winding space'' constructed in \cite{Berman:2014jsa} (see also the references therein), 
as solutions of the complete theory without simplifying assumptions. 

Further, it will be important to establish a formulation in which the background of an arbitrary defect brane 
can be equally described globally; 
that is, a manifestly $U$-duality covariant formulation of the supergravity theory. 
A promising approach in this direction is taken in DFT, 
which can globally describe both the usual brane (i.e.~NS5-brane) and the exotic brane (i.e.~$5^2_2$-brane), 
and can reproduce the standard supergravity action or the $\beta$-supergravity action as a special limit. 
Although DFT has already been generalized to incorporate the Ramond-Ramond fields \cite{Hohm:2011dv,Jeon:2012kd,Jeon:2012hp}, 
we cannot globally describe $U$-folds (such as the $\DD p_{7-p}$ background) in the framework of DFT, 
since the gauge symmetry of DFT does not include general $U$-dual transformations. 
Recently, several generalizations of DFT have been studied in various papers 
(see e.g.~\cite{Berman:2010is,Berman:2011kg,Berman:2011cg,Berman:2011jh,Berman:2012uy,Berman:2012vc,Park:2013gaj,Hohm:2013jma,Park:2014una}, 
and \cite{Hohm:2013pua,Hohm:2013vpa,Hohm:2013uia,Godazgar:2014nqa,Hohm:2014fxa,Blair:2014zba,Berman:2014hna,Musaev:2014lna} 
where the \emph{exceptional field theory} has been proposed and studied), 
which will be possible to describe all exotic-brane backgrounds globally. 
It will be interesting to derive the effective theories proposed in this paper 
as some special limits of these theories. 

It will be also important to investigate a reformulation of the effective worldvolume theory of exotic branes 
by using the newly introduced background fields $\bigl(\tilde{g}_{ij},\,\tilde{\phi},\,\cA^{i_1\cdots i_{7-p}}\bigr)$. 
More generally, it will be important to find a manifestly $U$-duality covariant formulation for 
the effective worldvolume theory of exotic branes. 

In our reformulation of the supergravity where a $p$-vector is regarded as a fundamental field, 
there naturally appears the background of an instanton that electrically couples to a $p$-vector. 
Depending on the type of the $p$-vector (i.e.~$\beta^{ij}$, $\gamma^{i_1\cdots i_{7-p}}$, or $\beta^{i_1\cdots i_6}$), 
the value of the on-shell action for the instanton background 
is expected to be proportional to $\gs^{\tilde{\alpha}}$ with $\tilde{\alpha}=0,1,2$. 
Since these instantons have not been studied well, 
it will be important to analyze their properties further. 
Since the $\II_1^{7-p}$ backgrounds \eqref{eq:gamma-instantons} are similar 
to the backgrounds of the D$p$-instantons (see \cite{Hull:1998vg} and references therein), 
it will be natural to expect that the instantons $\II_1^{7-p}$ are the exotic dual of the D$p$-instantons, 
and $\II^2_0$ and $\II^6_{2}$ are their $S$-dual objects.

\section*{Acknowledgments}

We acknowledge support by the National Research Foundation of Korea (NRF-MSIP) 
grants 2005-0093843, 2010-220-C00003 and 2012-K2A1A9055280. 
We would like to thank Yolanda Lozano and Soo-Jong Rey for helpful discussions. 

\appendix

\section{Field redefinitions and a derivation of the action}
\label{app:derivation}

In this appendix, we show that the action \eqref{eq:p-vector-action} can be rewritten as \eqref{eq:gamma-action-p} 
under the assumption \eqref{eq:assumptions}. 
Let us recall that, under a rescaling of the metric $\sfg_{ij}=\Exp{\sigma}G_{ij}$ in a $d$-dimensional space, 
the Ricci scalar transforms as
\begin{align}
 R=\Exp{\sigma}\,\bigl\{\sfR +(d-1)\,\sfg^{ij}\,\bar{\nabla}_i\partial_j\sigma
 -[(d-1)(d-2)/4]\,\sfg^{ij}\,\partial_i\sigma\,\partial_j\sigma\bigr\}\,,
\label{eq:R-formula}
\end{align}
where $\bar{\nabla}_i$ and $\sfR$ are the covariant derivative and the Ricci scalar associated with the new metric $\sfg_{ij}$. 
Using the formula, we can rewrite the action \eqref{eq:p-vector-action}
\begin{align}
\begin{split}
 2\kappa_{10}^2\,S_p 
 &= \int \biggl[\Exp{-2\phi}\,\bigl(*\,R + 4\,\rmd \phi\wedge *\,\rmd \phi\bigr)
\\
 &\qquad - \frac{\Delta^{-1}}{2\,(7-p)!}\, G_{\bfr_1\bfs_1}\cdots G_{\bfr_{7-p}\bfs_{7-p}}\,
           \rmd \gamma^{\bfr_1\cdots \bfr_{7-p}}\wedge *\,\rmd\gamma^{\bfs_1\cdots \bfs_{7-p}}\biggr] \,,
\end{split}
\end{align}
into the following form:
\begin{align}
\begin{split}
 2\kappa_{10}^2\,S_p &= 
    \int \rmd^{10}x\,\sqrt{\abs{\sfg}}\, \Exp{-2\phi-4\sigma}\,
 \bigl(R + 4\,\sfg^{ij}\,\partial_i\phi\,\partial_j \phi 
      +9\,\sfg^{ij}\,\bar{\nabla}_i\partial_j\sigma 
      -18\,\sfg^{ij}\,\partial_i\sigma\,\partial_j\sigma\bigr)
\\
 &\quad - \int\frac{\rmd^{10}x\sqrt{\abs{\sfg}}}{2\,(7-p)!}\,\Exp{-2\bvarphi}\,\sfg^{\bfa\bfb}\,\sfg_{\bfr_1\bfs_1}\cdots \sfg_{\bfr_{7-p}\bfs_{7-p}}\,
          \partial_\bfa \gamma^{\bfr_1\cdots \bfr_{7-p}}\, \partial_\bfb \gamma^{\bfs_1\cdots \bfs_{7-p}} \,,
\end{split}
\end{align}
where $\bvarphi$ is defined by $\Exp{\bvarphi}\equiv \Exp{\frac{11-p}{2}\,\sigma} \Delta^{1/2}$. 
In order to reduce the redundancy, 
let us consider the Einstein frame, $\phi=-2\sigma$, 
where the action takes the following form:
\begin{align}
\begin{split}
 2\kappa_{10}^2\,S_p &= 
    \int \rmd^{10}x\, \sqrt{\abs{\sfg}}\, \bigl(\sfR -2\,\sfg^{\bfa\bfb}\,\partial_\bfa\sigma\,\partial_\bfb\sigma\bigr) 
\\
 &\quad - \int\frac{\rmd^{10}x\sqrt{\abs{\sfg}}}{2\,(7-p)!}\,\Exp{-2\bvarphi}\,\sfg^{\bfa\bfb}\,\sfg_{\bfr_1\bfs_1}\cdots \sfg_{\bfr_{7-p}\bfs_{7-p}}\,
          \partial_\bfa \gamma^{\bfr_1\cdots \bfr_{7-p}}\, \partial_\bfb \gamma^{\bfs_1\cdots \bfs_{7-p}} \,,
\end{split}
\label{eq:S_p-temporary}
\end{align}
where the boundary term is dropped. 
Here, note that, under the ansatz \eqref{eq:assumptions}, the Ricci scalar $\sfR$ can be decomposed as
\begin{align}
 \sfR= \sfR^{(2)} +\frac{1}{4}\,\sfg^{\bfa\bfb}\,\partial_\bfa \sfg_{\bfp\bfq}\,\partial_\bfb \sfg^{\bfp\bfq} \,,
\end{align}
where $\sfR^{(2)}$ is the two-dimensional Ricci scalar associated with the metric $\sfg_{\bfa\bfb}$\,. 

Now, we make a further redefinition of the metric
\begin{align}
 \tilde{\sfg}_{\bfa\bfb}= \Exp{\tilde{\sigma}}\sfg_{\bfa\bfb}\,,\quad 
 \tilde{\sfg}_{\bfp\bfq}=\sfg_{\bfp\bfq} \,,
\end{align}
where $\tilde{\sigma}$ is a function to be specified below. 
Then, the formula \eqref{eq:R-formula} gives
\begin{align}
 \sfR^{(2)} =\Exp{\tilde{\sigma}}\,\bigl(\tilde{\sfR}^{(2)}+ \tilde{\sfg}^{\bfa\bfb}\,\tilde{\nabla}^{(2)}_\bfa\partial_\bfb\tilde{\sigma}\bigr) \,,
\end{align}
where $\tilde{\nabla}^{(2)}_\bfa$ and $\tilde{\sfR}^{(2)}$ are the covariant derivative 
and the Ricci scalar associated with the two-dimensional metric $\tilde{\sfg}_{\bfa\bfb}$.
That is, two Ricci scalars $\sfR$ and $\tilde{\sfR}$ (that is associated with the metric $\tilde{\sfg}_{ij}$) 
are related to each other through
\begin{align}
 \sfR= \Exp{\tilde{\sigma}}\,\bigl(\tilde{\sfR} + \tilde{\sfg}^{\bfa\bfb}\,\tilde{\nabla}^{(2)}_\bfa\,\partial_\bfb\tilde{\sigma}\bigr) \,.
\end{align}
Using this relation, we can rewrite the action \eqref{eq:S_p-temporary} as
\begin{align}
 2\kappa_{10}^2\,S_p &= 
    \int \rmd^{10}x\, \sqrt{\abs{\tilde{\sfg}}}\, \bigl(\tilde{\sfR} + \tilde{\sfg}^{\bfa\bfb}\,\tilde{\nabla}^{(2)}_\bfa\,\partial_\bfb\tilde{\sigma}
 -2\,\tilde{\sfg}^{\bfa\bfb}\,\partial_\bfa\sigma\,\partial_\bfb\sigma\bigr) 
\nn\\
 &\quad - \int\frac{\rmd^{10}x\sqrt{\abs{\tilde{\sfg}}}}{2\,(7-p)!}\,\Exp{-2\bvarphi}\,\tilde{\sfg}^{\bfa\bfb}\,
          \tilde{\sfg}_{\bfr_1\bfs_1}\cdots \tilde{\sfg}_{\bfr_{7-p}\bfs_{7-p}}\,
          \partial_\bfa \gamma^{\bfr_1\cdots \bfr_{7-p}}\, \partial_\bfb \gamma^{\bfs_1\cdots \bfs_{7-p}}
\nn\\
&= \int \rmd^{10}x\, \sqrt{\abs{\tilde{\sfg}}}\,
  \bigl(\tilde{\sfR} - \tilde{\sfg}^{\bfa\bfb}\,\partial_\bfa\bigl(\ln\!\sqrt{\abs{\det \tilde{\sfg}_{\bfp\bfq}}}\bigr)\,\partial_\bfb\tilde{\sigma}
        -2\,\tilde{\sfg}^{\bfa\bfb}\,\partial_\bfa\sigma\,\partial_\bfb\sigma\bigr) 
\nn\\
 &\quad - \int\frac{\rmd^{10}x\sqrt{\abs{\tilde{\sfg}}}}{2\,(7-p)!}\,\Exp{-2\bvarphi}\,\tilde{\sfg}^{\bfa\bfb}\,
          \tilde{\sfg}_{\bfr_1\bfs_1}\cdots \tilde{\sfg}_{\bfr_{7-p}\bfs_{7-p}}\,
          \partial_\bfa \gamma^{\bfr_1\cdots \bfr_{7-p}}\, \partial_\bfb \gamma^{\bfs_1\cdots \bfs_{7-p}} 
\nn\\
&= \int \rmd^{10}x\, \sqrt{\abs{\tilde{\sfg}}}\, 
   \Bigl[\tilde{\sfR} - \tilde{\sfg}^{\bfa\bfb}\,\partial_\bfa\Bigl(\bvarphi+ \frac{p-3}{2}\,\sigma\Bigr)\,\partial_\bfb\tilde{\sigma}
   -2\,\tilde{\sfg}^{\bfa\bfb}\,\partial_\bfa\sigma\,\partial_\bfb\sigma\Bigr] 
\nn\\
 &\quad - \int\frac{\rmd^{10}x\, \sqrt{\abs{\tilde{\sfg}}}}{2\,(7-p)!}\,\Exp{-2\bvarphi}\,\tilde{\sfg}^{\bfa\bfb}\,
          \tilde{\sfg}_{\bfr_1\bfs_1}\cdots \tilde{\sfg}_{\bfr_{7-p}\bfs_{7-p}}\,
          \partial_\bfa \gamma^{\bfr_1\cdots \bfr_{7-p}}\,\partial_\bfb\gamma^{\bfs_1\cdots \bfs_{7-p}} \,,
\end{align}
where we dropped a boundary term in the second equality 
and used 
$\sqrt{\abs{\det \tilde{\sfg}_{\bfp\bfq}}}=\Exp{4\sigma} \Delta^{1/2} =\Exp{\bvarphi+\frac{p-3}{2}\,\sigma}$ 
in the third equality. 

If we choose the function $\tilde{\sigma}$ as
\begin{align}
 \tilde{\sigma} = \frac{8}{(p-3)^2} \,\bvarphi - \frac{4}{p-3}\,\sigma \,,
\end{align}
the action becomes
\begin{align}
 2\kappa_{10}^2\,S_p 
 &= \int \rmd^{10}x\, \sqrt{\abs{\tilde{\sfg}}}\, \Bigl(\tilde{\sfR} -\frac{8}{(p-3)^2}\,\tilde{\sfg}^{\bfa\bfb}\,\partial_\bfa \bvarphi\,\partial_\bfb\bvarphi\Bigr) 
\nn\\
 &\quad - \int\frac{\rmd^{10}x\, \sqrt{\abs{\tilde{\sfg}}}}{2\,(7-p)!}\,\Exp{-2\bvarphi}\,
          \tilde{\sfg}^{\bfa\bfb}\,\tilde{\sfg}_{\bfr_1\bfs_1}\cdots \tilde{\sfg}_{\bfr_{7-p}\bfs_{7-p}}\,
          \partial_\bfa \gamma^{\bfr_1\cdots \bfr_{7-p}}\,\partial_\bfb\gamma^{\bfs_1\cdots \bfs_{7-p}}
\nn\\
 &= \int \rmd^{10}x\, \sqrt{\abs{\tilde{\sfg}}}\, \Bigl(\tilde{\sfR} -\frac{1}{2}\,\tilde{\sfg}^{\bfa\bfb}\,\partial_\bfa \tilde{\phi}\,\partial_\bfb\tilde{\phi}\Bigr)
\nn\\
 &\quad - \int\frac{\rmd^{10}x\, \sqrt{\abs{\tilde{\sfg}}}}{2\,(7-p)!}\,\Exp{-\frac{p-3}{2}\,\tilde{\phi}}\,
          \tilde{\sfg}^{\bfa\bfb}\,\tilde{\sfg}_{\bfr_1\bfs_1}\cdots \tilde{\sfg}_{\bfr_{7-p}\bfs_{7-p}}\,
          \partial_\bfa \gamma^{\bfr_1\cdots \bfr_{7-p}}\,\partial_\bfb\gamma^{\bfs_1\cdots \bfs_{7-p}} \,,
\end{align}
where we rescaled the dilaton; $\tilde{\phi}\equiv 4 \bvarphi/(p-3)$\,. 
Finally, defining the string-frame metric by $\tilde{g}_{ij}\equiv \Exp{\tilde{\phi}/2}\tilde{\sfg}_{ij}$, 
we obtain the action
\begin{align}
\begin{split}
 2\kappa_{10}^2\,S_p 
 &= \int \rmd^{10}x\, \sqrt{-\tilde{g}}\Exp{-2\tilde{\phi}}\, \bigl(\tilde{R} +4\,\tilde{g}^{\bfa\bfb}\,\partial_\bfa \tilde{\phi}\,\partial_\bfb\tilde{\phi}\bigr) 
\\
 &\quad - \int\frac{\rmd^{10}x\, \sqrt{-\tilde{g}}}{2\,(7-p)!}\,\Exp{-4\tilde{\phi}}\,
          \tilde{g}^{\bfa\bfb}\,\tilde{g}_{\bfr_1\bfs_1}\cdots \tilde{g}_{\bfr_{7-p}\bfs_{7-p}}\,
          \partial_\bfa \gamma^{\bfr_1\cdots \bfr_{7-p}}\,\partial_\bfb\gamma^{\bfs_1\cdots \bfs_{7-p}} \,,
\end{split}
\end{align}
which coincides with \eqref{eq:gamma-action-p} by allowing the bold indices to run over all spacetime directions. 
To summarize, we derived the action with the following redefinitions:
\begin{align}
\begin{split}
 (\tilde{g}_{ij}) 
 &= \bpm \Exp{\frac{4}{p-3}\,\phi} \Exp{\frac{p+1}{p-3}\,\eta} G_{\bfa\bfb} & 0\\
    0 & \Exp{\eta} G_{\bfp\bfq} \epm\,,
\quad 
 \Exp{2\tilde{\phi}} \equiv \Exp{2\phi+4\eta}\,,\quad 
 \Exp{\eta}\equiv \bigl(\Exp{-2\phi}\Delta^{1/2}\bigr)^{\frac{2}{p-3}} \,.
\end{split}
\end{align}

\end{document}